\documentclass[fleqn,usenatbib]{mnras}
\newcommand{\al}{\alpha}

\newcommand{\dl}{\delta}

\newcommand{\Lm}{\Lambda}

\newcommand{\f}{\frac}

\newcommand{\Msun}{$\mathrm{M}_{\sun}~$}

\makeatletter
\newenvironment{syslineq}[1][10]{%
  \mbox\bgroup\begin{math}\left\{%
  \begin{array}{*{#1}{cr}l}
}{%
  \end{array}%
  \right.\end{math}\egroup%
}
\makeatother

\newcommand{\citeg}[1]{\citep[e.g.,][]{#1}}
\usepackage{newtxtext,newtxmath}
\usepackage[T1]{fontenc}

\DeclareRobustCommand{\VAN}[3]{#2}
\let\VANthebibliography\thebibliography
\def\thebibliography{\DeclareRobustCommand{\VAN}[3]{##3}\VANthebibliography}

\usepackage{graphicx}	% Including figure files
\usepackage{here}
\title[The M31 stellar halo from HSC/\textit{NB515}]{The structure of the stellar halo of the Andromeda galaxy explored with the \textit{NB515} for Subaru/HSC. I.: New insights on the stellar halo up to 120 kpc}

\author[I. Ogami et al.]{
Itsuki Ogami,$^{1,2}$\thanks{E-mail: itsuki.ogami@grad.nao.ac.jp}
Mikito Tanaka$^{3}$,
Yutaka Komiyama$^{3}$,
Masashi Chiba$^{4}$,
Puragra Guhathakurta$^{5}$,
\newauthor
Evan N. Kirby$^{5}$,
Rosemary F. G. Wyse$^{7}$,
Carrie Filion$^{8}$,
Karoline M. Gilbert$^{9,10}$,
Ivanna Escala$^{11}$,
\newauthor
Masao Mori$^{12,13}$,
Takanobu Kirihara$^{14}$,
Masayuki Tanaka$^{1,2}$
Miho N. Ishigaki$^{1,2}$,
Kohei Hayashi$^{4,15}$,
\newauthor
Myung Gyoon Lee$^{16}$,
Sanjib Sharma$^{10}$,
Jason S. Kalirai$^{9}$,
and
Robert H. Lupton$^{17}$,
\\
$^{1}$The Graduate University for Advanced Studies (SOKENDAI), 2-21-1 Osawa, Mitaka, Tokyo 181-8588, Japan\\
$^{2}$National Astronomical Observatory of Japan, 2-21-1 Osawa, Mitaka, Tokyo 181-8588, Japan\\
$^{3}$Department of Advanced Sciences, Faculty of Science and Engineering, Hosei University, 3-7-2 Kajino-cho, Koganei, Tokyo 184-8584, Japan\\
$^{4}$Astronomical Institute, Tohoku University, Aoba-ku, Sendai, Miyagi 980-8578, Japan\\
$^{5}$Department of Astronomy and Astrophysics, University of California Santa Cruz, University of California Observatories, 1156 High Street, Santa Cruz, \\~~~CA 95064, USA\\
$^{6}$Department of Physics and Astronomy, University of Notre Dame, Notre Dame, IN 46556, USA\\
$^{7}$Department of Physics and Astronomy, Johns Hopkins University, Baltimore, MD 21218, USA\\
$^{8}$Center for Computational Astrophysics, Flatiron Institute, New York, NY 10010, USA\\
$^{9}$John Hopkins Applied Physics Laboratory, 11100 Johns Hopkins Road, Laurel, MD 20723, USA\\
$^{10}$Space Telescope Science Institute, 3700 San Martin Dr., Baltimore, MD 21218, USA\\
$^{11}$The Observatories of the Carnegie Institution for Science, 813 Santa Barbara Street, Pasadena, CA 91101, USA\\
$^{12}$Faculty of Pure and Applied Physics, University of Tsukuba, Tennodai 1-1-1, Tsukuba, Ibaraki, 305-8577, Japan\\
$^{13}$Center for Computational Sciences, University of Tsukuba, Tennodai 1-1-1, Tsukuba, Ibaraki, 305-8577, Japan\\
$^{14}$Kitami Institute of Technology, 165, Koen-cho, Kitami, Hokkaido 090-8507, Japan\\
$^{15}$National Institute of Technology, Sendai College, Natori, Miyagi 981-1239, Japan\\
$^{16}$Astronomy Program, Department of Physics and Astronomy, SNU Astronomy Research Center, Seoul National University, 1 Gwanak-ro, Gwanak-gu,\\~~~~Seoul 08826, Republic of Korea\\
$^{17}$Department of Astrophysical Sciences, Princeton University, 4 Ivy Lane, Princeton, NJ 08544, USA
}

\date{Accepted XXX. Received YYY; in original form ZZZ}

\pubyear{2023}

\begin{document}
\label{firstpage}
\pagerange{\pageref{firstpage}--\pageref{lastpage}}
\maketitle

\begin{abstract}
We analyse the M31 halo and its substructure within a projected radius of 120 kpc using a combination of Subaru/HSC \textit{NB515} and CFHT/MegaCam \textit{g}- \& \textit{i}-bands. We succeed in separating M31’s halo stars from foreground contamination with $\sim$ 90 \% accuracy by using the surface gravity sensitive \textit{NB515} filter. Based on the selected M31 halo stars, we discover three new substructures, which associate with the Giant Southern Stream (GSS) based on their photometric metallicity estimates. We also produce the distance and photometric metallicity estimates for the known substructures. While these quantities for the GSS are reproduced in our study, we find that the North-Western stream shows a steeper distance gradient than found in an earlier study, suggesting that it is likely to have formed in an orbit closer to the Milky Way. For two streams in the eastern halo (Stream C and D), we identify distance gradients that had not been resolved. Finally, we investigate the global halo photometric metallicity distribution and surface brightness profile using the \textit{NB515}-selected halo stars. We find that the surface brightness of the metal-poor and metal-rich halo populations, and the all population can be fitted to a power-law profile with an index of $\al=-1.65\pm0.02$, $-2.82\pm0.01$, and $-2.44\pm0.01$, respectively. In contrast to the relative smoothness of the halo profile, its photometric metallicity distribution appears to be spatially non-uniform with nonmonotonic trends with radius, suggesting that the halo population had insufficient time to dynamically homogenize the accreted populations.
\end{abstract}

\begin{keywords}
galaxies: individual: M31 -- Galaxies, galaxies: haloes -- Galaxies, Local Group
\end{keywords}

%%%%%%%%%%%%%%%%%%%%%%%%%%%%%%%%%%%%%%%%%%%%%%%%%%

%%%%%%%%%%%%%%%%% BODY OF PAPER %%%%%%%%%%%%%%%%%%

\section{Introduction}\label{section:intro}
Understanding how galaxies have formed and evolved to what we see today is one of the most critical issues in astronomy. \citet{searle1978} proposed that large galaxies such as the Milky Way (MW) and the Andromeda Galaxy (M31) have been formed by the accretions of small stellar systems over a long time, based on the lack of spatial dependence in the metallicity of globular clusters. This galaxy formation scenario is consistent with the Lambda-dominated cold dark matter ($\Lm$CDM) theory. The hierarchical structure formation scenario based on $\Lm$CDM model is also suitable for describing large-scale structures beyond the galactic scale ($\sim$ 1 Mpc) and is therefore widely accepted. This model predicts that the accreted stellar systems are disrupted by the tidal forces of the host galaxy, leaving behind tidal streams \citep{bullock2005,johnston2008,font2011}.

Stellar haloes are a key to understanding galaxy formation because they retain past chemodynamical information. Many of the halo stars found in the MW have low metallicities and high-velocity dispersions, which differ from the properties of disc stars \citeg{feltzing2013}. Analyses of the kinematics and chemical abundances in Galactic halo stars have uncovered that the Galactic halo consists of at least two distinct components, a flattened inner halo with a higher mean metallicity at $R_{\rm gal} < 10~{\rm kpc}$ and a spherical outer halo with a lower mean metallicity at $R_{\rm gal} > 30~{\rm kpc}$ \citep{carollo2007,carollo2010,conroy2019}. Thanks to the Gaia satellite mission, it has become evident that a large fraction of the field halo stars in the solar neighborhood exhibit large orbital eccentricities and lower $[\mathrm{\al/Fe}]$ ratios when compared to the majority of stars with similar metallicities. These results are currently interpreted as a stellar population that was accreted into the Milky Way approximately 10 billion years ago \citep{helmi2018,belokurov2018,gallart2019}. In addition to these halo stellar populations, observations with the Sloan Digital Sky Survey \citep[SDSS;][]{york2000}, the Gaia satellite \citep{gaiacollaboration2016a}, and others have discovered over 70 substructures \citeg{ibata1994,belokurov2006,belokurov2018,mateu2017,mateu2018,shipp2018}. These substructures are evidence that stellar haloes are formed by the accretion of small stellar systems. Furthermore, numerical simulations of the orbits of accreted dwarf galaxies based on the $\Lm$CDM theory revealed that past accretion events have caused dynamical heating of the Galactic disc \citeg{villalobos2008,cooper2010,grand2017} and that in-situ star formation was predicted in the Galactic halo \citeg{zolotov2009,zolotov2010}, such in-situ formation was observationally detected \citep[e.g.,][]{gallart2019,belokurov2022}. Thus, a detailed study of Galactic stellar haloes is an important clue to understanding galaxy formation.

Similar to the MW, the resolved disc and halo stars in our neighbor, M31, have also been observed and studied in detail. In particular, its low disc inclination of $12\degr5$ \citep{simien1978} allows M31 to provide a comprehensive view of the halo. Therefore, M31 is an excellent system for studying galaxy formation.

Many photometric and spectroscopic observations in the M31 halo have been conducted, including large-scale surveys such as Pan-Andromeda Archaeological Survey \citep[PAndAS;][]{mcconnachie2009,mcconnachie2018} with the Canada France Hawaii Telescope (CFHT)/MegaCam and the Spectroscopic and Photometric Landscape of Andromeda’s Stellar Halo \citep[SPLASH;][]{gilbert2009} survey with the Keck II/DEep Imaging Multi-Object Spectrograph (DEIMOS). The results of these observations show that there are some similarities to the MW halo, but various differences have also been found. Like the Galactic halo, the surface brightness of the M31 outer halo follows a power-law profile, and many stars have low metallicities ($[\mathrm{Fe/H}]<-1.5$) \citep{guhathakurta2005,kalirai2006,ibata2014,gilbert2014,gilbert2020}. On the other hand, the inner halo of M31 has been found to have higher metallicities ($[\mathrm{Fe/H}]\sim-0.6$) and younger ages ($\sim 8 \mathrm{Gyr}$) than the Galactic inner halo \citeg{brown2003,brown2006,gilbert2014}. In addition to the differences in the metallicities, the masses of stellar halos of the MW and M31 are very different; the stellar halo of the MW has $\sim 10^9~M_{\odot}$ \citep{deason2019}, and the accreted masses of M31 was expected to be $>1-2\times 10^{10}~M_{\odot}$ \citep{dsouza2018}.

Substructures have also been discovered in the M31 halo similar to the Galactic halo. Starting with the discovery of the Giant Southern Stream (GSS) by \citet{ibata2001}, more than 10 substructures have been discovered in the eastern and western parts of the M31 disc and in the northwestern and southwestern parts of the halo \citeg{ibata2001,ferguson2002,ferguson2005,ibata2007,mcconnachie2009,richardson2011}.

As in the Galaxy, these observations provide many opportunities for simulations. In particular, GSS formation scenarios have been put forward through numerical simulations: this substructure can be formed from radial merger with a dwarf galaxy of masses of $\sim 10^9$ \Msun \citeg{fardal2006,fardal2007,mori2008,kirihara2014,kirihara2017a}, or in a major, gas-rich merger \citeg{dsouza2018,hammer2018}. This accreted event may also produce shell structures in the east and west of the M31 disc, and these shell structures have indeed been shown to be associated with GSS by spectroscopic observations \citep{escala2021,dey2023}. Comparing the observations of the M31 halo with those of the Galactic halo, it is suggested that disc galaxies such as the Galaxy and M31 have undergone different accretion and evolutionary paths, suggesting that individual stellar haloes differ in their star formation history and dwarf galaxy accretion times \citep[e.g.,][]{bullock2005,cooper2010,merritt2016,monachesi2016a,harmsen2017a,harmsen2023}. Thus, detailed observations of the M31 stellar halo are of great importance in gaining new insights into formation scenarios for the formation of galactic haloes.

However, since M31 is located at low galactic latitude, many foreground stars of the Galaxy make it difficult to distinguish M31 halo stars from such contaminations. Therefore, it is difficult to observe low surface brightness structures of the M31 halo. These Galactic foreground stars are expected to increase exponentially with decreasing galactic latitude, accounting for up to 70 \% of all stars per deg$^2$ \citep{martin2013}. Therefore, although substructures have been discovered in M31, little progress has been made in estimating the distributions of their distances and line-of-sight velocities from observations and in simulating the origin of these substructures, except for the GSS and the stellar stream in the northwestern part of the M31 halo \citep[NW Stream;][]{mcconnachie2009}.

To overcome this situation, a narrow-band filter (\textit{NB515}) was developed for the Subaru Telescope/Hyper Suprime-Cam (HSC). With \textit{NB515}, red giant branch stars (RGBs) in the M31 halo can be efficiently separated from foreground dwarf stars in the MW based on differences in surface gravity between these stellar populations. In fact, for the NW Stream located at low galactic latitude ($\mathrm{b} \sim -15\degr$), observations with \textit{NB515} succeeded to extract only M31 halo stars and estimate the distance to the stream \citep{komiyama2018a}. From this result, it has become possible to constrain the accretion trajectory that reproduces the NW Stream \citep{kirihara2017a}. Using {\it DDO51}, a intermediate filter for the Mosaic Camera on the Kitt peak NationalObservatory 4m Mayall telescope , RGB stars in the periphery of M31 were extracted, followed by spectroscopic observations using Keck/DEIMOS \citep[e.g.,][]{gilbert2006}. These observations detected RGB stars at distances up to 175 kpc in the M31 halo \citep{gilbert2012}. Therefore, the use of narrow-band filter such as \textit{NB515} for Subaru/HSC is powerful for understanding the physical properties of M31 substructures.

In this paper, we report observations and analyses of the M31 halo using the Subaru/HSC \textit{NB515}. The HSC is a prime focus camera on the Subaru Telescope with a field of view diameter of 1.5 deg \citep{miyazaki2012,miyazaki2018,furusawa2018,komiyama2018}. We focus on the eight substructures already discovered in M31 and the properties of the halo from the data obtained with \textit{NB515}. From this analysis, we demonstrate the effectiveness of \textit{NB515} by comparing the results with and without \textit{NB515}, and also show the powerfulness of \textit{NB515} for future HSC analysis into deep broadband data. The paper is organized as follows. In Section \ref{section:data_and_method}, we describe our observational and archival data and the method for data analysis of the M31 halo. The results of the distance estimation of the M31 substructure, the metallicity distribution, and the surface brightness profile of the halo are given in Section \ref{section:results}, and the discussion for these results are presented in Section \ref{section:discussion}. Finally, Section \ref{section:conclusions} concludes this paper.

\section{Data and method}\label{section:data_and_method}

\subsection{Data}\label{subsection:data}

We use Pan-Andromeda Archaeological Survey \citep[PAndAS;][]{mcconnachie2009,mcconnachie2018} \textit{g}- and \textit{i}-bands data combined with Subaru/HSC {\it NB515} data as explained below to investigate the nature of the M31 stellar halo. The PAndAS data provides the wide and deep observational data for the M31/M33 stellar halo, so many studies have been conducted using the PAndAS data to understand the structure of the M31 stellar halo and to identify dwarf galaxies and stellar substructures in the M31 halo \citeg{mcconnachie2009,martin2013,ibata2014}. However, most of these studies assumed the statistical model to eliminate foreground contaminations \citep{martin2013}. In this study, to reduce the uncertainties associated with such a model we use the \textit{NB515} filter for Subaru/HSC in combination with these PAndAS/\textit{g}- and \textit{i}-band data to select the M31 halo stars (see Section \ref{subsection:nb515} and \ref{subsection:selection_RGB}) and to achieve the deep and homogeneous analysis for the M31 halo without contaminations.

\begin{figure}
  \begin{center}
   \includegraphics[width=\columnwidth]{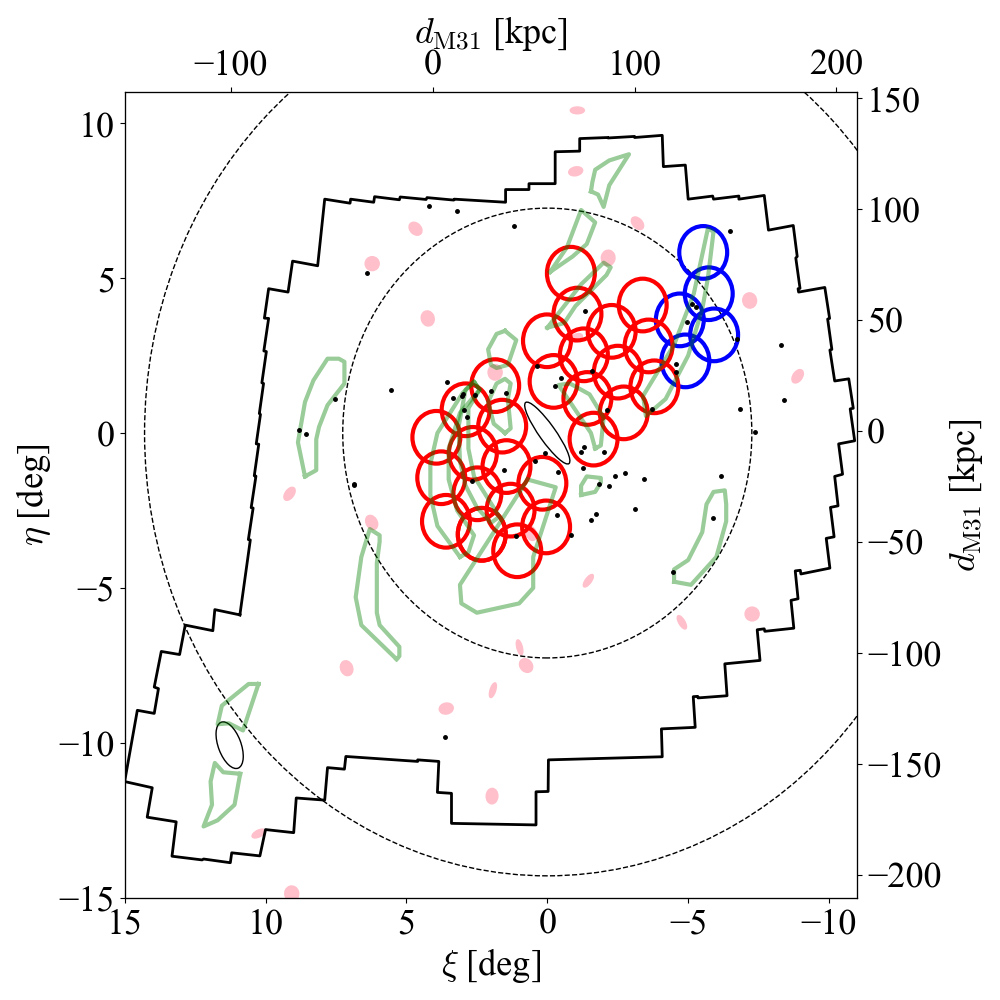} 
  \end{center}
 \caption{HSC survey fields (red and blue circles; fields coloured by red were observed in 2019 and fields coloured by blue were observed in 2015) showing tangent plane centered on M31 with the positions of major stellar substructures (green polygons), dwarf galaxies (pink ellipces), and globular clusters (black dots). The black ellipses corresponds to the M31 and M33 discs. The outer black frame indicates the PAndAS region.}\label{figure:observed_map}
\end{figure}

The M31 halo was observed in 2015 and 2019 using the Subaru/HSC. Figure \ref{figure:observed_map} shows the observed regions in this study, where red and blue circles are the fields observed in 2019 (28 fields) and 2015 (5 fields), respectively \citep{komiyama2018}. These observed fields occupy about 50 deg$^2$ of the M31 halo, and each field was observed multiple times to avoid CCD gaps and cosmic ray effects. The integration time varies from 960 to 1920 seconds, with most fields having integration times of 960 seconds, and its seeing is $0\farcs50 - 1\farcs1$, with a mean value of $0\farcs8$ and a standard deviation of $0\farcs2$. Details of HSC observations are shown in Table \ref{table:obs_condition}.

\begin{table*}
 \caption{The details of observations in HSC. Field names with the prefix "PFS\_FIELD\_" show the red circles, and field names with the prefix "M31\_" show the blue circles in Figure \ref{figure:observed_map}}
 \label{table:obs_condition}
 \begin{tabular}{@{}l@{\hspace*{43pt}}l@{\hspace*{43pt}}l@{\hspace*{43pt}}l@{\hspace*{43pt}}l@{\hspace*{43pt}}l@{}}
       \hline
       Field & $\al (\mathrm{J}2000)$ & $\dl (\mathrm{J}2000)$ & Date (mm/dd/yy) & Exposure Time & Seeing FWHM \\
       \hline
       PFS\_FIELD\_1 & $00^{\mathrm{h}}34^{\mathrm{m}}53\fs9$ & $+42\degr22\arcmin57\farcs0$ & 09/30/2019 & $240\times4$ s &  $0\farcs65 - 0\farcs85$\\
       PFS\_FIELD\_2 & $00^{\mathrm{h}}28^{\mathrm{m}}55\fs9$ & $+43\degr11\arcmin01\farcs0$ & 09/30/2019 & $240\times4$ s &  $0\farcs65 - 0\farcs75$\\
       PFS\_FIELD\_3 & $00^{\mathrm{h}}22^{\mathrm{m}}48\fs9$ & $+43\degr57\arcmin54\farcs0$ & 09/30/2019 & $240\times4$ s &  $0\farcs60 - 0\farcs75$\\
       PFS\_FIELD\_6 & $00^{\mathrm{h}}41^{\mathrm{m}}42\fs9$ & $+42\degr54\arcmin25\farcs0$ & 09/30/2019 & $240\times4$ s &  $0\farcs60 - 0\farcs70$\\
       PFS\_FIELD\_7 & $00^{\mathrm{h}}35^{\mathrm{m}}46\fs9$ & $+43\degr43\arcmin47\farcs0$ & 09/30/2019 & $240\times4$ s &  $0\farcs55 - 0\farcs65$\\
       PFS\_FIELD\_8 & $00^{\mathrm{h}}29^{\mathrm{m}}41\fs9$ & $+44\degr32\arcmin01\farcs0$ & 09/30/2019 & $240\times4$  s &  $0\farcs55 - 0\farcs75$\\
       PFS\_FIELD\_9 & $00^{\mathrm{h}}23^{\mathrm{m}}25\fs9$ & $+45\degr19\arcmin01\farcs0$ & 09/30/2019 & $240\times4$ s &  $0\farcs65 - 0\farcs75$\\
       PFS\_FIELD\_11 & $00^{\mathrm{h}}42^{\mathrm{m}}44\fs9$ & $+44\degr15\arcmin02\farcs0$ & 09/30/2019 & $240\times4$ s &  $0\farcs60 - 0\farcs75$\\
       PFS\_FIELD\_12 & $00^{\mathrm{h}}36^{\mathrm{m}}41\fs9$ & $+45\degr04\arcmin35\farcs0$ & 09/30/2019 & $240\times4$ s &  $0\farcs65 - 0\farcs65$\\
       PFS\_FIELD\_13 & $00^{\mathrm{h}}37^{\mathrm{m}}39\fs9$ & $+46\degr25\arcmin20\farcs1$ & 09/30/2019 & $240\times4$ s &  $0\farcs60 - 0\farcs85$\\
       PFS\_FIELD\_14 & $00^{\mathrm{h}}34^{\mathrm{m}}03\fs9$ & $+41\degr02\arcmin05\farcs0$ & 09/30/2019 & $240\times4$  s&  $0\farcs85 - 0\farcs95$\\
       PFS\_FIELD\_15 & $00^{\mathrm{h}}28^{\mathrm{m}}12\fs9$ & $+41\degr50\arcmin00\farcs1$ & 09/30/2019 & $240\times4$ s &  $0\farcs85 - 0\farcs85$\\
       PFS\_FIELD\_16 & $00^{\mathrm{h}}22^{\mathrm{m}}12\fs9$ & $+42\degr36\arcmin45\farcs0$ & 09/30/2019 & $240\times4$ s &  $0\farcs75 - 0\farcs90$\\
       PFS\_FIELD\_19 & $00^{\mathrm{h}}50^{\mathrm{m}}17\fs9$ & $+40\degr07\arcmin25\farcs0$ & 09/30/2019 & $240\times4$ s &  $0\farcs75 - 0\farcs90$\\
       PFS\_FIELD\_20 & $00^{\mathrm{h}}55^{\mathrm{m}}43\fs9$ & $+39\degr15\arcmin27\farcs0$ & 09/30/2019 & $240\times4$ s &  $0\farcs70 - 0\farcs85$\\
       PFS\_FIELD\_21 & $00^{\mathrm{h}}01^{\mathrm{m}}01\fs9$ & $+38\degr22\arcmin34\farcs1$ & 09/30/2019 & $240\times4$ s &  $0\farcs75 - 0\farcs80$\\
       PFS\_FIELD\_22 & $00^{\mathrm{h}}51^{\mathrm{m}}27\fs9$ & $+41\degr27\arcmin45\farcs0$ & 09/30/2019 & $240\times4$ s &  $0\farcs75 - 0\farcs75$\\
       PFS\_FIELD\_23 & $00^{\mathrm{h}}56^{\mathrm{m}}58\fs9$ & $+40\degr35\arcmin34\farcs1$ & 09/30/2019 & $240\times4$ s &  $0\farcs75 - 0\farcs75$\\
       PFS\_FIELD\_24 & $00^{\mathrm{h}}02^{\mathrm{m}}21\fs9$ & $+39\degr42\arcmin28\farcs0$ & 09/30/2019 & $240\times4$ s &  $0\farcs75 - 0\farcs85$\\
       PFS\_FIELD\_25 & $00^{\mathrm{h}}52^{\mathrm{m}}40\fs9$ & $+42\degr48\arcmin01\farcs0$ & 09/30/2019 & $240\times4$ s &  $0\farcs75 - 0\farcs95$\\
       PFS\_FIELD\_26 & $00^{\mathrm{h}}58^{\mathrm{m}}17\fs9$ & $+41\degr55\arcmin37\farcs1$ & 09/30/2019 & $240\times5$ s &  $0\farcs95 - 1\farcs1$\\
       PFS\_FIELD\_27 & $00^{\mathrm{h}}03^{\mathrm{m}}45\fs9$ & $+41\degr02\arcmin17\farcs0$ & 10/01/2019 & $240\times4$ s &  $0\farcs80 - 0\farcs95$\\
       PFS\_FIELD\_28 & $00^{\mathrm{h}}43^{\mathrm{m}}42\fs9$ & $+39\degr37\arcmin51\farcs0$ & 10/01/2019 & $240\times4$ s &  $0\farcs70 - 0\farcs85$\\
       PFS\_FIELD\_29 & $00^{\mathrm{h}}49^{\mathrm{m}}11\fs9$ & $+38\degr47\arcmin03\farcs1$ & 10/01/2019 & $240\times4$ s &  $0\farcs65 - 0\farcs80$\\
       PFS\_FIELD\_30 & $00^{\mathrm{h}}54^{\mathrm{m}}31\fs9$ & $+37\degr55\arcmin17\farcs1$ & 10/01/2019 & $240\times4$ s &  $0\farcs80 - 0\farcs95$\\
       PFS\_FIELD\_31 & $00^{\mathrm{h}}37^{\mathrm{m}}12\fs9$ & $+39\degr06\arcmin52\farcs1$ & 10/01/2019 & $240\times5$ s &  $0\farcs90 - 1\farcs0$\\
       PFS\_FIELD\_32 & $00^{\mathrm{h}}42^{\mathrm{m}}43\fs9$ & $+38\degr17\arcmin16\farcs0$ & 10/01/2019 & $240\times8$ s &  $0\farcs95 - 1\farcs1$\\
       PFS\_FIELD\_33 & $00^{\mathrm{h}}48^{\mathrm{m}}05\fs9$ & $+37\degr26\arcmin39\farcs1$ & 10/01/2019 & $240\times4$ s &  $0\farcs85 - 0\farcs95$\\
       M31\_003 & $00^{\mathrm{h}}16^{\mathrm{m}}31\fs7$ & $+44\degr43\arcmin30\farcs0$ & 10/12/2015 & $240\times4$ s & $0\farcs65 - 0\farcs75$\\
       M31\_004 & $00^{\mathrm{h}}10^{\mathrm{m}}04\fs7$ & $+45\degr27\arcmin47\farcs0$ & 10/12/2015 & $240\times4$ s & $0\farcs55 - 0\farcs65$\\
       M31\_009 & $00^{\mathrm{h}}10^{\mathrm{m}}24\fs5$ & $+46\degr49\arcmin07\farcs0$ & 10/12/2015 & $240\times4$ s & $0\farcs55 - 0\farcs65$\\
       M31\_022 & $00^{\mathrm{h}}16^{\mathrm{m}}04\fs2$ & $+43\degr22\arcmin15\farcs0$ & 10/09/2015 & $240\times4$ s & $0\farcs85 - 0\farcs95$\\
       M31\_023 & $00^{\mathrm{h}}09^{\mathrm{m}}45\fs8$ & $+44\degr06\arcmin28\farcs0$ & 10/09/2015 & $240\times4$ s & $0\farcs85 - 0\farcs95$\\
       \hline
 \end{tabular}
\end{table*}

The area that the PAndAS data covers, which can be accessed from the Canadian Astronomy Data Centre (CADC)\footnote[1]{https://www.cadc-ccda.hia-iha.nrc-cnrc.gc.ca/en/community/pandas/\\query.html} is shown in Figure \ref{figure:observed_map} within the outer black frame. PAndAS covers over $400$ deg$^2$ of the M31 halo and M33 as well, using the \textit{g}- and \textit{i}-bands, respectively. The integration time is 500 and 1160 s for \textit{g}- and \textit{i}-bands, and the data are available from the tip of the red-giant branch (TRGB) of the M31 population to at least 3 magnitude fainter than TRGB. In the HSC observations, we performed dithering to compensate for the large and small gaps between the CCDs. Similar to the HSC observations, PAndAS also performed dithering by shifting the field of view of each shot by approximately $10\arcsec$. Although this could compensate for the small gap between each CCD, it was not enough to compensate for the large gap between the CCDs. Therefore, geometric gaps can be seen in the final spatial distribution (e.g., Figure \ref{figure:RGBprob_CMD}) in this study.

\subsection{Reduction and Photometry}\label{subsection:reduction_photometry}

The data observed by HSC are processed using the hscPipe 6.7 \citep{bosch2018}. The hscPipe is based on the Large Synoptic Survey Telescope \citep[LSST;][]{ivezic2008} pipeline and optimized for HSC. In the hscPipe, data reduction is performed on individual CCD, including de-bias, dark subtraction, flat fielding, sky subtraction, and cosmic ray removal. After the reduction of individual CCD, the coordinates and flux scales of each CCD are calibrated, each frame is co-added, and finally, the source detection and photometry are performed to construct the source catalogue. In the hscPipe, it uses Pan-STARRS1 \citep[PS1;][]{schlafly2012,tonry2012,magnier2013,flewelling2020} for the photometric and astrometric calibration.

% The PAndAS data is processed by the Elixir pipeline \citep{magnier2004}, and the Cambridge Astronomical Survey Unit (CASU) pipeline \citep{irwin2004} is used for detection, photometry, and astrometry. The Elixir performs de-biasing, flat-fielding, and an estimate of the photometric zero point on the observed CCDs. After the processing, the data are calibrated using the CASU pipeline for photometric and astrometric calibration.

The final multi-colour catalogue used in this study is constructed by cross-matching the HSC/\textit{NB515} catalogue with the PAndAS catalogue under the condition that the position of each star is $< 0\farcs5$. In this study, we use only the sources that are determined to be stellar-like in both the hscPipe and PAndAS catalogues. In other words, we consider as stellar sources that are determined to have $\mathrm{\texttt{extendedness}} = 0$, which is indicators of successful star-galaxy classification at depths from ${\it i}\sim 23~{\rm mag}$ \citep{aihara2018a} in hscPipe and have a probability of being a point source within $1\sigma$ or $2\sigma$ in both \textit{g}- and \textit{i}-bands of the PAndAS catalog.

Based on the dust maps of \citet{schlegel1998}, we apply Galactic extinction correction for our catalogue. The extinction-corrected magnitudes for each band are
\begin{align}
\textit{g}_0 &= \textit{g} - 3.793\times E(B-V) \notag \\
\textit{i}_0 &= \textit{i} - 2.086\times E(B-V)\\
\textit{NB515}_0 &= \textit{NB515} - 2.862\times E(B-V) \notag
\end{align}
where the subscript zero means the extinction-corrected magnitude. The coefficient of the extinction-corrected magnitude of \textit{NB515} is obtained by multiplying the interstellar absorption curve of \citet{fitzpatrick1999} with $R=+3.1$ by the SED of a G-type star ($\mathrm{T_{eff}}=7000$ K, $\log{\mathrm{Z}}=-1$, $\log{g}=4.5$) and integrating it using the \textit{NB515} response curve. For $g_0$ and $i_0$, we use the coefficient from PAndAS studies \citep[e.g.,][]{ibata2014,mcconnachie2018}.

\begin{figure}
  \includegraphics[width=\columnwidth]
  {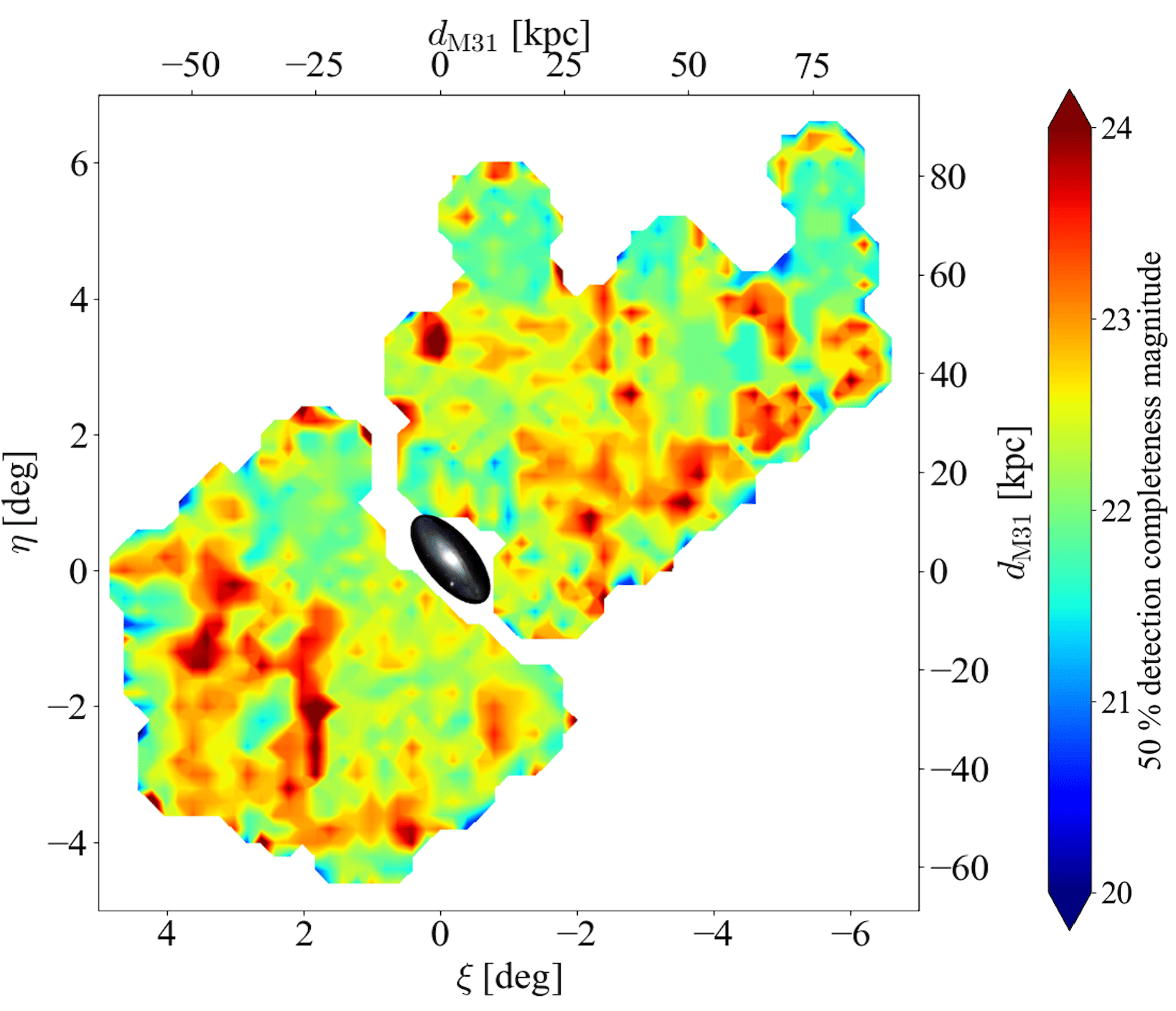}
 \caption{Apparent magnitudes at 50 \% completeness for \textit{NB515} for each $0.1 \times 0.1~{\rm degree}^2$ region. The colour image of M31 observed by HSC is inserted in the centre (credit: NAOJ).}
 \label{figure:completeness}
\end{figure}

The detection completeness varies from region to region depending on the signal-to-noise ratio and the crowding of objects. We evaluated the detection completeness of \textit{NB515} by embedding and detecting artificial stars in the images. To perform this test, we developed a Python module \texttt{injectStar.py}\footnote[2]{https://github.com/itsukiogami/injectStar}. Details of this module and the artificial star test are described in Appendix \ref{section:completeness}. Figure \ref{figure:completeness} shows the apparent magnitudes corresponding to the 50 \% completeness of \textit{NB515} for each $0.1 \times 0.1~{\rm degree}^2$ region. The average magnitude of the 50 \% completeness in the \textit{NB515} band is $23.21$ mag. In this map, the region where seeing is slightly worse (the center coordinate of the tangential plane are $(3,1)$ for ${\rm PFS}\_{\rm FIELD}\_26$, $(-1,-2)$ for ${\rm PFS}\_{\rm FIELD}\_31$, $(0,-3)$ for ${\rm PFS}\_{\rm FIELD}\_32$), the 50\% detection completeness magnitude is the almost same as the other regions. Detection completeness for the PAndAS survey was calculated and modeled by \citet{martin2016}. They showed the completeness $\eta(m)$ in equation (\ref{equation:completeness}).
\begin{align}
 \eta(m) = \f{A}{1+\exp{\left(\f{m-m_{50}}{\rho}\right)}} \label{equation:completeness}
\end{align}
where, $m$ is the apparent magnitude of the \textit{g}- and \textit{i}-bands, $m_{50}$ is the apparent magnitude of 50\% completeness, $A$ and $\rho$ are the constant values. In the {\it g}-band, $m_{50} = 24.88$, $A = 0.94$, and $\rho = 0.65$, while in the {\it i}-band, $m_{50} = 23.88$, $A = 0.93$, and $\rho = 0.74$. In this study, the completeness model is assumed to be uniform in the overall region. By fitting the results of the artificial star test (the ratio of the number of detected artificial stars to the number of injected artificial stars) to equation (\ref{equation:completeness}), we obtain $m_{50} = 23.21$, $A = 0.99$, and $\rho = 0.30$ for \textit{NB515}.

\subsection{Colour-Magnitude Diagram}\label{subsection:cmd}

\begin{figure}
 \includegraphics[width=\columnwidth]
 {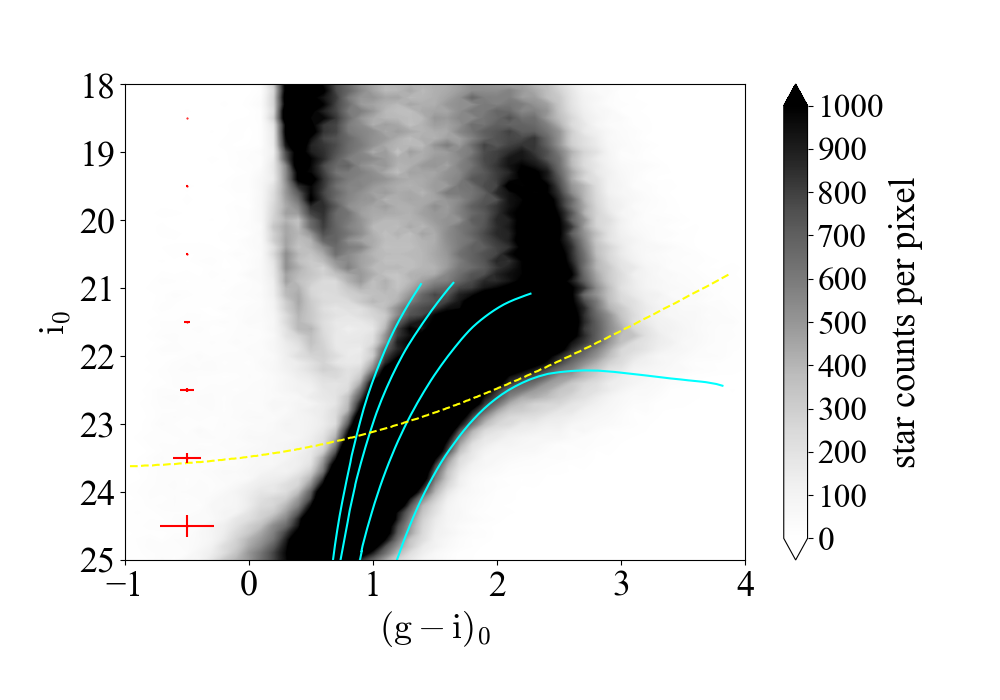}
 \vspace{-10mm}
 \caption{The colour-magnitude diagram (CMD) for all analysed regions shown as the Hess diagram with bin size $0.01~\mathrm{mag} \times 0.01~\mathrm{mag}$. The cyan curves are Dartmouth isochrones, for the [$\al$/Fe] ratio of 0.0 and the age of 13 Gyr, with $[\mathrm{Fe/H}] = 0.00, -0.70, -1.50, -2.00$ from right to left. The yellow dashed line shows the $50~\%$ detection completeness, and the red crosses indicate the typical photometric errors.}
 \label{figure:CMD}
\end{figure}

Figure \ref{figure:CMD} shows the de-reddened colour-magnitude diagram (CMD) shown as the Hess diagram in the entire region analysed in this work, and the Dartmouth isochrones \citep{dotter2008} with $[\al/\mathrm{Fe}] = 0.0$, age of 13 Gyr, and different metallicities of $[\mathrm{Fe/H}]=0.00, -0.70, -1.50, -2.00$ which are scaled to the distance of M31 \citep[776 kpc;][]{dalcanton2012}. The yellow dashed line shows the $50\%$ completeness, and the red crosses are the average photometric errors at $(g - i)_0 = 1.0$. From this figure, we can identify various stellar populations. The red giant branch (RGB) stars in the M31 halo, the target of this study, are distributed from $((g - i)_0, i_0) \sim (2.0, 21)$ to $\sim(0.5, 25)$. The region of $18 \loa i_0 \loa 22$ at $(g - i)_0 \sim 2.2$ is occupied by the main sequence (dwarf) stars in the foreground Galactic disc, while the region of $18 \loa i_0 \loa 22$ at $(g - i)_0 \sim 0.5$ shows the main sequence turn-off (MSTO) stars in the Galactic halo. These stellar populations indicate that the RGB stars in M31 overlap with the foreground Galactic stars in the CMD. Previous PAndAS works have used data with $i_0<23$ to account for the detection completeness and photometric errors \citeg{martin2013, bate2014}, so we also use the stars with $i_0<23$ for the analysis.

\subsection{\textit{NB515} and colour-colour diagram}\label{subsection:nb515}

Figure \ref{figure:CMD} also shows that the distribution of the M31 RGB stars is overlapping with that of the Galactic dwarf stars in the CMD, making it difficult to distinguish these stellar populations from only the $(g_0, i_0)$ information. For this purpose, we use the Subaru/HSC narrow-band filter \textit{NB515}, which is sensitive to the surface gravity of a star: the intensity of spectral absorption in the MgH band at 521 nm and in the Mgb triplet at 518 nm strongly depend on the stellar surface gravity \citep{ohman1934,thackeray1939}. Based on this property, \citet{majewski2000} proposed a method to separate RGB stars from dwarf stars using \textit{DDO51} which is an intermediate-band filter with a central wavelength of 515 nm and bandwidth (FWHM) of 12.3 nm \citep{clark1979}. Since the MgH+Mgb absorptions are covered by \textit{DDO51}, there is a difference in the \textit{DDO51} magnitude between the RGB and dwarf owing to different surface gravities: the colour index $(\textit{M} - \textit{DDO51})$, where \textit{M} is a broadband magnitude in the Washington system, provides an index of the surface gravity \citep{majewski2000}.

\begin{figure}
 \includegraphics[width=\columnwidth]
 {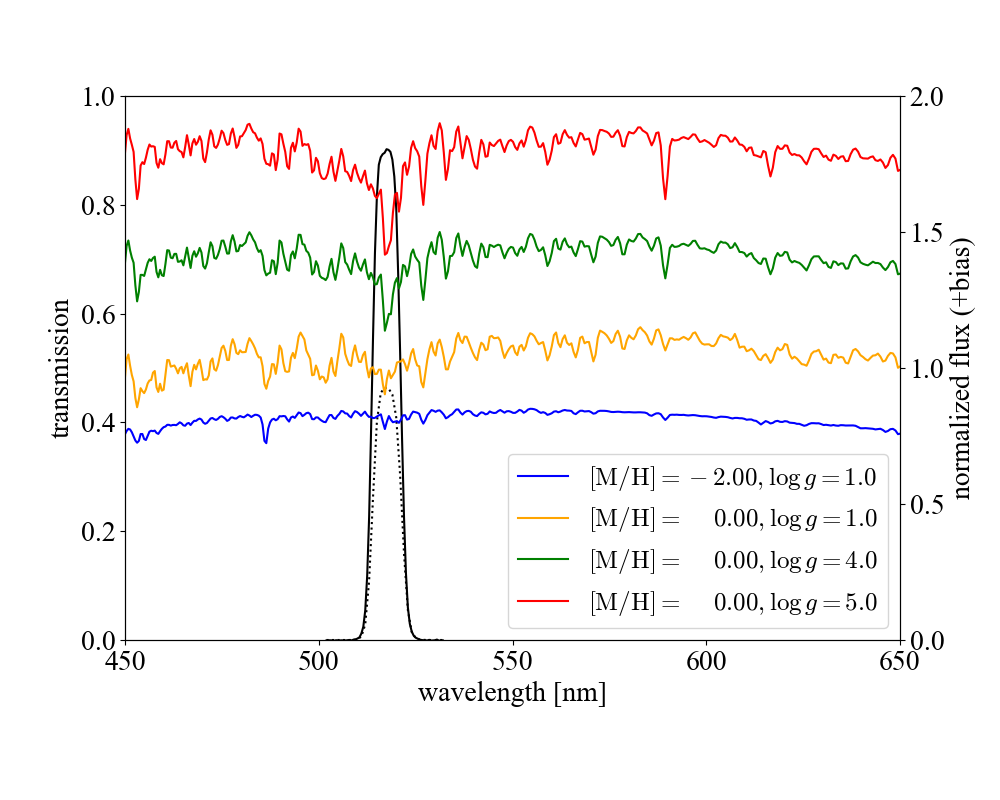}
 \vspace{-10mm}
 \caption{Response curves of the HSC/\textit{NB515} filter used in this study. The solid black line shows the response curves for the filter itself, and the dotted line shows the total response on the ground. The spectra like dwarf stars are red and green solid lines, and the spectra like RGB stars are orange and blue from the BOSZ library. Note that red, green, and orange lines are shifted upwards for clarity. The shape of the absorption lines in RGB stars is almost the same whether they are metal-poor (blue) or metal-rich (orange).}
 \label{figure:nb515}
\end{figure}

Following this property of \textit{DDO51}, \textit{NB515} for HSC is made with a central wavelength of 515 nm and an FWHM of 7.7 nm by our group (PI: M. Chiba). In Figure \ref{figure:nb515}, the solid black line shows the sensitivity curve of \textit{NB515} only, and the dotted black line shows the sensitivity curve of \textit{NB515} multiplied by the atmospheric transmittance, the reflectance of the primary mirror, the transmittance of the prime focus optics, the quantum efficiency of the CCD, and the transmittance of the dewar window. Figure \ref{figure:nb515} also shows stellar model spectra with $\mathrm{T_{eff}} = 5000$ K, $[\mathrm{C/M}] = 0.00$, and $[\al/\mathrm{M}] = 0.25$ BOSZ stellar model spectra\footnote[3]{https://archive.stsci.edu/prepds/bosz/} \citep{bohlin2017}. The red and green lines show dwarf stars with $[\mathrm{Fe/H}]$ of $0.00$ and $\log{g}$ of 4 and 5, respectively, while the orange and blue lines show giant stars with $[\mathrm{Fe/H}]$ of $0.00$ and $-2.00$, respectively, and $\log{g} = 1$. In Figure \ref{figure:nb515}, the absorption lines that overlap with the sensitivity curve of \textit{NB515} correspond to MgH+Mgb, and it can be confirmed that these lines for RGB stars (orange and blue) are weaker than those for dwarf stars (red and green). Following the method proposed by \citet{majewski2000}, we use three photometric bands (\textit{g, i, NB515}), which cover a similar wavelength range as those photometric bands used in \citet{majewski2000}, to extract M31 RGB stars on the colour-colour diagram.

\begin{figure}
 \includegraphics[width=\columnwidth]
 {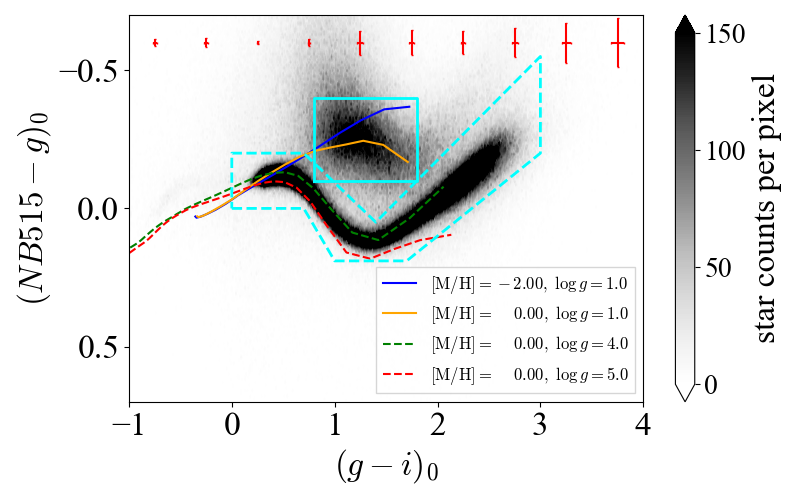}
 \vspace{-5mm}
 \caption{$(\textit{g} - \textit{i})_0-(\textit{NB515} - \textit{g})_0$  diagram for all stellar sources in this study, shown as the Hess diagram with bin size $0.01 \mathrm{mag} \times 0.01\mathrm{mag}$. Blue and orange solid lines indicate the RGB-like model locus with $[\mathrm{M/H}] = -2.00$, $\log{g} = 1.0$, and with $[\mathrm{M/H}] = 0.00$, $\log{g} = 1.0$. Green and red dashed lines show the dwarf-like model locus with $[\mathrm{M/H}] = 0.00$, $\log {g} = 4.0$, and with $[\mathrm{M/H}] = 0.00$, $\log{g} = 5.00$. The cyan rectangle and dashed polygon indicate the distribution where  RGB/dwarf stars are easy to concentrate. The red crosses indicate the typical photometric errors.}
 \label{figure:2CD}
\end{figure}

Figure \ref{figure:2CD} shows the colour-colour diagram of the entire analysis region and the theoretical curves calculated by BOSZ model spectra with the properties of the dwarf and RGB stars. The blue and orange solid lines are the RGB-like model spectra with $[\mathrm{M/H}] = -2.00$, $\log{g} = 1.0$, and with $[\mathrm{M/H}] = 0.00$, $\log{g} = 1.0$, respectively. The green and red dashed lines correspond to the dwarf-like model spectra with $[\mathrm{M/H}] = 0.00$, $\log{g} = 4.0$, and with $[\mathrm{M/H}] = 0.00$, $\log{g} = 5.0$, respectively. It is clearly seen that the distribution of RGB stars (enclosed by a solid cyan rectangle) is different from that of dwarf stars (enclosed by a dashed cyan polygon). Note that the metal-rich giant (orange solid line in Figure \ref{figure:2CD}) intersects the distribution of many dwarf stars (a cyan dashed polygon) with $(\textit{g}-\textit{i})_0 > 2$. Therefore, our \textit{NB515}-based selection can not work well on the metal-rich side, so to account for this, we cut these red stars in our analysis (metallicity profiles and surface brightness profiles in the halo of M31; Section \ref{section:results}).

\subsection{Selection of the M31 RGB Stars}\label{subsection:selection_RGB}

In this subsection, we derive the dwarf probability (RGB probability) based on narrow-band information, $p_{\mathrm{dwarf, \textit{NB}}}$ ($p_{\mathrm{RGB, \textit{NB}}}$) and dwarf probability (RGB probability) based on galactic latitude information $p_{\mathrm{dwarf, \textit{lat}}}$ ($p_{\mathrm{RGB, \textit{lat}}}$) for each star. From the combination of these probabilities, we calculate the probability of being a RGB star in M31, $p_{\mathrm{M31}}$, for each star. 

\begin{figure*}
  \includegraphics[width=2\columnwidth]
  {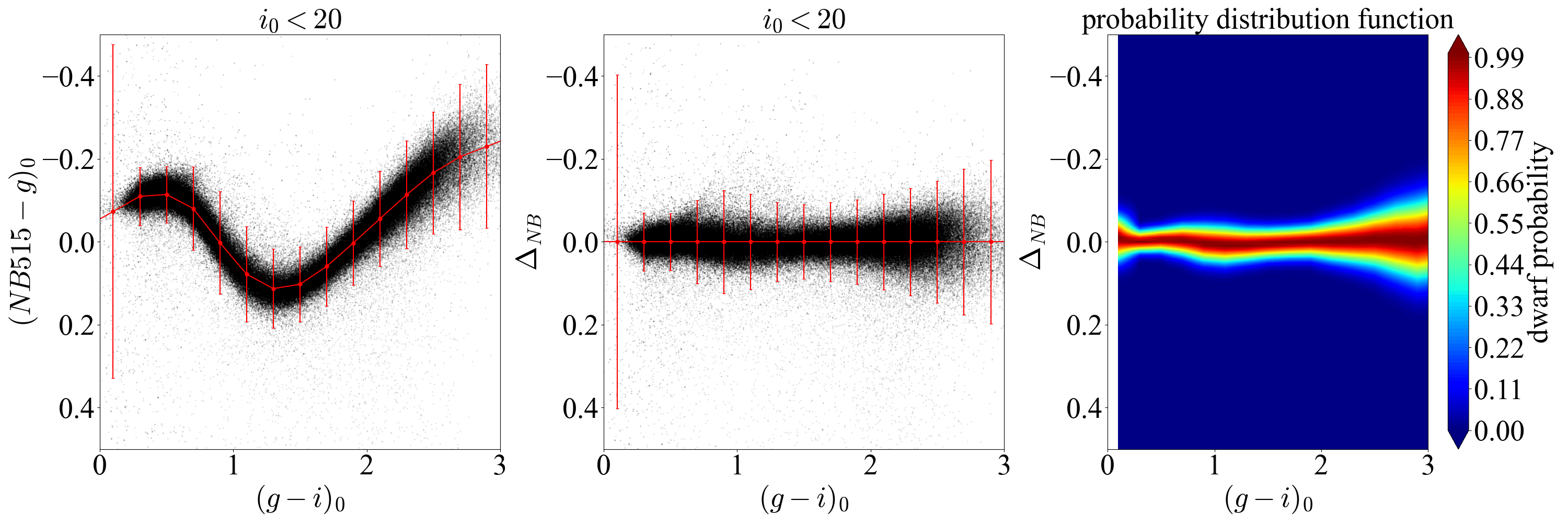}
 \vspace{-10pt}
 \caption{Left: The colour-colour diagram of stars with $i_0<20$. The red line and error bars show the Dwarf Ridge Line and 3 times standard deviations. Middle: The corrected colour-colour diagram. The red line and error bars show the Dwarf Ridge Line and 3 times standard deviations. Right: The dwarf probability distribution function on the colour-colour space.}
 \label{figure:2CD_dwarf}
\end{figure*}

First of all, we assume that the distance to disc of M31 is 776 kpc \citep{dalcanton2012} and all M31 RGB stars are located within a virial radius of $\sim$ 260 kpc \citep{seigar2008}. Then, while the M31 RGB stars are distributed mostly at the fainter side of $i_0 \gtrsim 21$ mag of the CMD, many foreground Galactic dwarf stars are available at the brighter side, in $i_0 < 20$ mag, where there exist no M31 RGBs. The left panel of Figure \ref{figure:2CD_dwarf} shows the colour-colour diagram of stars with $i_0 < 20$, which is regarded as the colour-colour diagram of only foreground dwarf stars. We bin the data between $0.0 \leq (\textit{g}-\textit{i})_0 \leq 3.0$ in $(\textit{g}-\textit{i})_0$ bins of width 0.2 mag. In each bin, we compute the mean and standard deviation of $(\textit{NB515}-\textit{g})_0$. By interpolation of these mean values using the \texttt{Python/SciPy} package, we construct the function that represents the distribution of dwarf stars (Dwarf Ridge Line; $f_{\mathrm{dwarf}}((g-i)_0)$). The left panel of Figure \ref{figure:2CD_dwarf} shows Dwarf Ridge Line $f_{\mathrm{dwarf}}((g-i)_0)$ as a solid red line, where the red dots and error bars depict the mean values and 3 times standard deviations, respectively, in each bin. 

To easily derive the $p_{\mathrm{dwarf},\textit{NB}}$, we introduce $\Delta_{\textit{NB}} \equiv (\textit{NB515}-\textit{g})_0-f_\mathrm{dwarf}((\textit{g}-\textit{i})_0)$, which makes the mean of the distribution of foreground dwarf stars a straight horizontal line on the diagram. The middle panel of Figure \ref{figure:2CD_dwarf} shows the $(\textit{g}-\textit{i})_0$-$\Delta_{\textit{NB}}$ diagram, where the vertical axis of the left panel of Figure \ref{figure:2CD_dwarf} is replaced by $\Delta_{\textit{NB}}$. We assume that the probability distribution of the dwarf sequence at a given $(g - i)_0$ is a Gaussian distribution with mean $\mu= 0$, and standard deviation $\sigma= \sigma_{(\textit{NB515}-\textit{g})_0}$ in $\Delta_{\textit{NB}}$ direction, where the standard deviation $\sigma_{(\textit{NB515}-\textit{g})_0}$ is calculated for each bin of 0.2 mag. Note that the photometric errors of dwarf stars ($\textit{i}_0<20$) are sufficiently small (mean value is $< 0.01$) that $\sigma_{(\textit{NB515}-\textit{g})_0}$ does not include photometric error.

Performing the linear interpolation of these normal distributions, we construct the probability distribution $P_{\mathrm{dwarf}}$ in which individual stars are likely to be the dwarf star on the $(\textit{g}-\textit{i})_0$-$\Delta_{\textit{NB}}$ diagram. The right panel of Figure \ref{figure:2CD_dwarf} shows this distribution.

Based on this right panel and the photometric errors of individual stars, we calculate the dwarf probability $p_{\mathrm{dwarf},\textit{NB}}$ for each star. First, we assume that each star is distributed as a two-dimensional Gaussian distribution based on the photometric errors in colour $\sigma_{(g-i)_0}$ and $\sigma_{(\textit{NB515}-\textit{g})}$. Then, the distribution function $\mathcal{N}((\textit{g}-\textit{i})_0,\Delta_{\textit{NB}})|(g - i)_{0,n}, \Delta_{\textit{NB
},n})$ of the \textit{n}-th star with photometric errors of $(\sigma_{(\textit{g}-\textit{i})_{0,n}}, \sigma_{\Delta_{\textit{NB},n}})$ at $((\textit{g}-\textit{i})_{0,n}, \Delta_{\textit{NB},n})$ in the colour-colour diagram is expressed as
\begin{align}
    &\mathcal{N}((\textit{g}-\textit{i})_0,\Delta_{\textit{NB}}|(\textit{g}-\textit{i})_{0,n},\Delta_{\textit{NB},n}) \notag \\
    &= \f{1}{2\pi\sqrt{\det{V}}} \exp{\left[ -\f{1}{2}\left(
    \begin{bmatrix}
        (g-i)_0\\ \Delta_{\textit{NB}}\\
    \end{bmatrix}
    -\mu \right) V^{-1} \left(
    \begin{bmatrix}(g-i)_0\\
        \Delta_{\textit{NB}}\\
    \end{bmatrix}
    -\mu \right)\right]} \label{eq_pdf}
\end{align}
where,
\begin{align}
    \mu = 
    \begin{bmatrix}
        (g-i)_{0,n}\\
        \Delta_{\textit{NB},n}\\
    \end{bmatrix}
    , \ V = 
    \begin{bmatrix}
        \sigma_{(g-i)_{0,n}}^2 & {\rm Cov}((g-i)_0,\Delta_{NB}) \\
        {\rm Cov}((g-i)_0,\Delta_{NB}) & \sigma_{\Delta_{\textit{NB},n}}^2 \\
    \end{bmatrix} . \notag
\end{align}
, and ${\rm Cov}((g-i)_0,\Delta_{NB})$ is the covariance between $(g-i)_0$ and $\Delta_{NB}$. It is noted that when ${\rm Cov}((g-i)_0,\Delta_{NB})$ is larger than $\sigma_{(g-i)_{0,n}}^2$ or $\sigma_{\Delta_{\textit{NB},n}}^2$, $V$ is no longer a positive semidefinite matrix and we cannot calculate the bivariate normal distribution (Equation \ref{eq_pdf}). Therefore, when ${\rm Cov}((g-i)_0,\Delta_{NB})$ is greater than $\sigma_{(g-i)_{0,n}}^2$ or $\sigma_{\Delta_{\textit{NB},n}}^2$, ${\rm Cov}((g-i)_0,\Delta_{NB})$ is assumed to be 0, because when the photometric error is small, it is expected that the distribution of individual stars on the color-color diagram will not differ significantly.

The dwarf probability $p_{\mathrm{dwarf},\textit{NB}}$ for each star is calculated by integrating the product of the probability distribution of dwarf stars $P_{\mathrm{dwarf}}$ shown in the right panel of Figure \ref{figure:2CD_dwarf} and the distribution of individual stars shown in Equation (\ref{eq_pdf}) as follows
\begin{align}
    p_{\mathrm{dwarf},\textit{NB}} = \int_{(g-i)_0} \int_{\Delta_{\textit{NB}}} P_{\mathrm{dwarf}} \times \mathcal{N} \ d(g-i)_0 \ d\Delta_{\textit{NB}}, \label{equation:NBDwarfLikelihood}
\end{align}
and normalizing these calculated values. The RGB probability $p_{\mathrm{RGB},\textit{NB}}$ of each star is calculated to be $p_{\mathrm{RGB},\textit{NB}} = 1 - p_{\mathrm{dwarf},\textit{NB}}$, 
assuming that the stars distributed in the colour-colour diagram are only Galactic dwarf stars and M31 RGB stars. 

From the colour-colour diagram (see, Figure \ref{figure:2CD} and Figure \ref{figure:2CD_dwarf}), it is evident that dwarf stars generally have $\Delta_{\textit{NB}}>0$, while RGB stars have $\Delta_{\textit{NB}} < 0$. Therefore, we take into account this effect by multiplying $p_{\mathrm{RGB,\textit{NB}}}$ with the step function $f_{\mathrm{step}}$,
\begin{align}
    f_{\mathrm{step}}(\Delta_{\textit{NB}}) = 
\begin{syslineq}
\begin{aligned}
        1 ~, \mathrm{for}~\Delta_{\textit{NB}}<0 \\
        0 ~, \mathrm{for}~\Delta_{\textit{NB}}>0
\end{aligned}
\end{syslineq}\label{eq_step}
\end{align}
to make the value of $p_{\mathrm{RGB},\textit{NB}}$ smaller when $\Delta_{\textit{NB}} > 0$. 

The number of foreground stars is expected to change over our $\sim 50$ deg$^2$ field of view as a function of latitude \citep[see e.g.,][]{tanaka2010,martin2013}. We incorporate this spatial effect into our probabilities by computing $p_{\mathrm{RGB},lat}$ as follows. First, in the same way as \citet{tanaka2010},  we fit the spatial distribution of foreground-like stars with $i_0<20$ with a function of galactic latitude, \textit{b}, 
\begin{align}
f(\textit{b}|A,B,C) = A\times \exp{\left[ B\times \textit{|b|} + C \right]},
\label{eq_Lat_model}
\end{align}
where \textit{A}, \textit{B} and \textit{C} are the constant values. In this function, the returned value (dwarf probability of galactic latitude) is 1 at $\textit{b}=0$. Then, based on the spatial information for each star, we calculate the dwarf probability $p_{\mathrm{dwarf},\textit{lat}}$ of each star. The RGB probability $p_{\mathrm{RGB},\textit{lat}}$ based on the latitude is calculated as $p_{\mathrm{RGB}, lat} = 1-p_{\mathrm{dwarf}, lat}$ under the assumption that there are only RGB stars and dwarf stars as stellar populations in this analysis. Finally, based on the calculated RGB probabilities of each star based on its \textit{NB515} and galactic latitude, the probability of being an RGB star $p_{\mathrm{M31}}$ is calculated according to the following equation
\begin{align}
p_{\mathrm{M31}} = \frac{p_{\mathrm{RGB},\textit{NB}} \times p_{\mathrm{RGB},\textit{lat}}}{p_{\mathrm{RGB},\textit{NB}} \times p_{\mathrm{RGB},\textit{lat}} + p_{\mathrm{dwarf},\textit{NB}} \times p_{\mathrm{dwarf},\textit{lat}}}.
\label{eq_M31_RGBProbability}
\end{align}
where the denominator is the normalization constant.

\begin{figure*}
  \includegraphics[width=2\columnwidth]
  {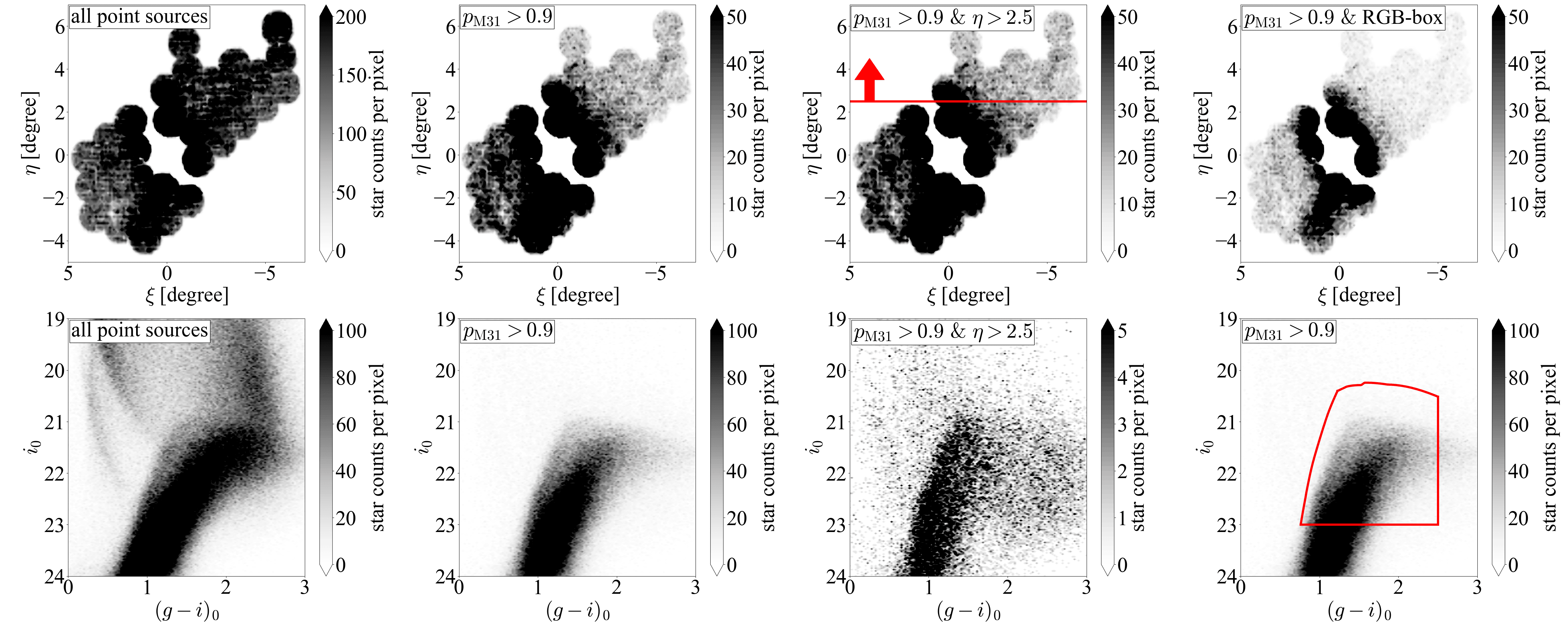} 
 \caption{Top: Spatial distributions of all point sources, stars with $p_{\mathrm{M31}}>0.9$, stars with $p_{\mathrm{M31}}>0.9$ and $\eta > 2.5$, and \textit{NB515}-based selected RGB stars (NRGB) from left to right. Bottom: Colour-magnitude diagrams of each population. In the rightmost panel, red lines show the edge of the RGB-box.}
 \label{figure:RGBprob_CMD}
\end{figure*}

Figure \ref{figure:RGBprob_CMD} shows the spatial distribution (top panels) and CMD (bottom panels) for all stars (leftmost panels) and those with M31 RGB probability $p_{\mathrm{M31}} > 0.9$ (inner-left panels). It is clear that the CMD for the stars with high $p_{\mathrm{M31}}$ show a more prominent RGB sequence than those without the selection by $p_{\mathrm{M31}}$. In the spatial distribution, although the number of objects decreases when the selection by $p_{\mathrm{M31}}$ is applied, we can confirm the already discovered substructures in the M31 halo \citep[e.g. GSS, Stream C, and Stream D][]{ibata2001,ferguson2002,ibata2007}. Therefore, by using the RGB probability $p_{\mathrm{M31}}$, the clear structure of the M31 halo is made available without the influence of contamination from the Galactic foreground dwarf stars. 

Using $p_{\mathrm{M31}}$, it is possible to extract a large number of M31 RGB stars. The inner-right bottom panel in Figure \ref{figure:RGBprob_CMD} shows the CMD for the stars with high $p_{\mathrm{M31}}$ and low galactic latitude ($\eta > 2.5$ on M31-centric coordinates). In this panel, we confirm that there are stars at $((\textit{g}-\textit{i})_0) \sim (2.5, 22)$, which look like remaining Galactic dwarf-like stars. As seen Figure \ref{figure:2CD}, it is difficult to separate the dwarf stars from the RGB stars at $(g-i)>2.5$. This causes some (red) dwarf stars to have high $p_{\mathrm{RGB}}$. It should be noted, however, that the contrast of the inner-right bottom panel in Figure \ref{figure:RGBprob_CMD} is different from the other CMDs, so we expect that the number of the remaining dwarf stars is small. In Section \ref{subsection:accuracy}, we verify the accuracy of the M31 RGB probability $p_{\mathrm{M31}}$ calculated in this study, confirm that the accuracy of $p_{\mathrm{M31}}$ is $\sim 90\%$.

In order to remove these remaining dwarf stars with high $p_{\mathrm{M31}}$ and investigate the pure properties of the M31 stellar halo, we further set a RGB-box to select the most likely RGBs on the CMD based on isochrones, where the age is 13 Gyr, $[\al/\mathrm{Fe}] = 0.0, Y = 0.245 + 1.5Z$, and $[\mathrm{Fe/H}] = -2.5$ to $+0.5$. To achieve this, we assume isochrones to be distributed at a distance of $785\pm100$ kpc, assuming the spread of the M31 stellar halo. The fainter boundary of $i_0$ of the RGB-box is set to $i_0 = 23$ by considering the completeness. The reddest of $(g-i)_0$ is set to $(g-i)_0 = 2.5$, considering that completeness decreases on the red side and that RGB stars overlap dwarf stars on the red side. The CMD in the rightmost bottom panel of Figure \ref{figure:RGBprob_CMD} shows the RGB-box as the red box. Hereafter, the objects with $p_{\mathrm{RGB}} > 0.9$ in this RGB-box are called NRGB stars (basically \textit{NB515}-based selected RGB stars). The NRGB stars are used to construct the photometric metallicity distribution described in Section \ref{subsubsection:MD} and the surface brightness profile described in Section \ref{subsubsection:SBprofile}.

\subsection{Algorithm for Distance Estimation}\label{subsection:algorithm}

In metal-poor stellar systems in which individual stars can be resolve, the brightness of the TRGB can be used as a distance indicator. Starting with \citet{lee1993}, the absolute \textit{i}-band magnitude of TRGB was found to vary by only 0.1 mag in stellar systems with $[\mathrm{Fe/H}] < -0.7$ \citeg{lee1993,bellazzini2005,rizzi2007}. This feature leads to a sharp drop at magnitudes brighter than the apparent magnitude of TRGB for stellar systems with low metallicity and a single stellar population. The luminosity function of such RGB stars with apparent magnitudes, \textit{m}, can be modeled as
\begin{align}
\begin{syslineq}
\begin{aligned}
        \Phi(m) &= 10^{a\left(m-m_{\mathrm{TRGB}}\right)} + b &, m>m_{\mathrm{TRGB}} \\
        \Phi(m) &= b &, m\leq m_{\mathrm{TRGB}}.
\end{aligned}\label{eq_LF}
\end{syslineq}
\end{align}
, where \textit{a} and \textit{b} are constants. The value of parameter {\it a} should originally be estimated using the initial mass function, but since it has been treated as a free parameter in RGB distance estimation\citeg{makarov2006,conn2011,tollerud2016}, so it is regarded as a free parameter in this study as well.

Such distance estimation methods are effective for stellar systems with low metallicity and a single population. However, for stellar systems with metal-rich and/or a wide range of metallicity, the TRGB brightness is not constant and no clear step-like distribution appears in the luminosity function, even for systems with a sufficient number of objects. Therefore, for stellar systems with metal-rich and/or a wide metallicity dispersion, it is not possible to fit with Equation (\ref{eq_LF}). To overcome this problem, a distance estimation method is developed by constructing a model CMD based on the RGB isochrones on the CMD and fitting the model to the observed data \citep{conn2016}. Some substructures that are analysed in this study include stellar systems that are known to have metal-rich and a wide metallicity dispersion. Therefore, the method developed by \citet{conn2016} was optimized and applied to the data in this analysis to estimate the distance. In this section, we briefly introduce the method used in \citet{conn2016} to construct the model CMD and the distance estimation method, as well as the optimization of the method to our analysed data.

Following the procedure of \citet{conn2016}, we perform the model fitting and distance estimation, applying likelihood calculations and the Markov Chain Monte Carlo method (MCMC) to the observed data. To construct the colour-magnitude model, we prepare the isochrones with the metallicity of $-2.5 \leq [\mathrm{Fe/H}] \leq 0.5$ and the age of 10 Gyr. The helium mass ratio is assumed to be $Y = 0.245 + 1.5Z$, and the $\al$-element ratio is assumed to be $[\al/\mathrm{Fe}] = 0.0$. In constructing the model, the distribution of the number of RGB stars on each isochrone is obtained and then the distribution is interpolated. The distribution of the number of RGB stars on each isochrone is constructed using the above equation (\ref{eq_LF}) and the following equation (\ref{eq_MD}). First, the number of RGB stars at a given magnitude $m$ is assumed to vary according to Equation (\ref{eq_LF}). In this equation, $m_{\mathrm{TRGB}}$, $a$, and \textit{b} are the parameters of the model. $m_{\mathrm{TRGB}}$ is the magnitude of TRGB, $a$ is the slope of the luminosity function, \textit{b} is the model of the foreground/background objects. The parameter $b$ is a constant, and $m$ is the apparent magnitude in the isochrone. In the Dartmouth isochrones, the available isochrone dataset is given by the absolute magnitude $M_{\mathrm{RGB}}$. Therefore, the absolute magnitude $M_{\mathrm{RGB}}$ needs to be converted to apparent magnitude $m_{\mathrm{RGB}}$. Here, we use the distance to the stellar system in kpc, $d$, and $d$ is a parameter to be estimated during model fitting. In the observed stellar system, the metallicity distribution is not uniform metallicity distribution. Therefore, it is necessary to reflect the gradient of the number of objects according to the metallicity distribution in the model CMD of RGB stars. The contribution of an isochrone with a certain metallicity $[\mathrm{Fe/H}]_{\mathrm{iso}}$ to the model CMD is given by 
\begin{align}
    W_{\mathrm{iso}} = \exp{\left( -\f{([\mathrm{Fe/H}]_{\mathrm{iso}}-[\mathrm{Fe/H}]_0)^2}{2\times \omega_{\mathrm{RGB}}^2} \right)} \label{eq_MD}
\end{align}
assuming that the metallicity distribution of the stellar system is the Gaussian distribution. The parameters $[\mathrm{Fe/H}]_0$ and $\omega_{\mathrm{RGB}}$ are the mean metallicity and standard deviation of the metallicity of the stellar system. Multiplying equations (\ref{eq_LF}) and (\ref{eq_MD}) determines the distribution of the number of RGB stars on each isochrone. Therefore, the change in the number of RGB stars in an isochrone with a certain metallicity $[\mathrm{Fe/H}]_{\mathrm{iso}}$ becomes
\begin{align}
    \Phi(m)\times W_{\mathrm{iso}} = 
\begin{syslineq}
\begin{aligned}
        & \left[ 10^{a\left(m-m_{\mathrm{TRGB}}\right)} + b \right] \times\\
        & \exp{\left( -\f{([\mathrm{Fe/H}]_{\mathrm{iso}}-[\mathrm{Fe/H}]_0)^2}{2\times \omega_{\mathrm{RGB}}^2}\right)} \\
        & ~~~~,\mathrm{for}~~m>m_{\mathrm{TRGB}} \\
        & b \\
        & ~~~~,\mathrm{for}~~m\leq m_{\mathrm{TRGB}}.
\end{aligned}
\end{syslineq}\label{eq_RGB}
\end{align}

\begin{figure}
 \includegraphics[width=\columnwidth]
 {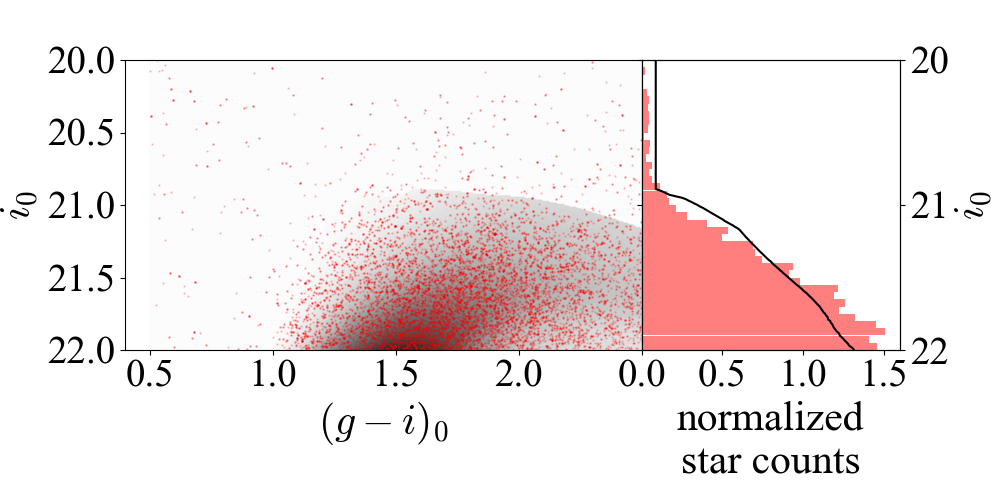}
 \vspace{-20pt}
 \caption{Model CMD and luminosity function of the area near the centre of M31 in the GSS (in Section \ref{subsection:substructures}, we call this region "GSS1") constructed by applying the distance estimation method in this study. In the left panel, the gray shaded region is the normalized model CMD, and red dots are observed data with $p_{\mathrm{M31}}>0.9$. In the right panel, a black solid line is the normalized model luminosity function derived from summing up the model CMD, and a red-shaded histogram is the normalized observed luminosity function.}
 \label{figure:modelCMD}
\end{figure}

A model CMD is constructed by three-dimensional linear interpolation of equation (\ref{eq_RGB}). Figure \ref{figure:modelCMD} shows the model around the centre of M31 in the GSS created by applying the distance estimation method. The red dots are the observed data, and it can be confirmed that the model (black shaded) reproduces the observed RGB stars (the distribution from $((g - i)_0 \sim 2, i_0 \sim 21)$ to $((g - i)_0 \sim 1.5, i_0 \sim 22))$. \citet{conn2016} use the model constructed above to estimate distances by calculating the likelihood $\mathcal{L}_{\mathrm{CMD},n}$ of n-th observed data and applying it to MCMC.

In addition to the above, this study considers the detection completeness and RGB probability $p_{\mathrm{M31}}$ for the calculation of the likelihood. For the n-th star, we calculate the detection completeness ($\eta(\textit{g}_{0,n}), \eta(\textit{i}_{0,n}), \eta(\textit{NB515}_{0,n})$) from equation (\ref{equation:completeness}) and the probability $p_{\mathrm{M31},n}$ (see, Sectioon \ref{subsection:selection_RGB}). We assume that the contribution of the n-th data to the CMD of the substructure is weighted by $p_{\mathrm{M31}}$ and the inverse of the detection completeness. Under this assumption, we can derive the likelihood by directly multiplying these values to the likelihood $\mathcal{L}_{\mathrm{CMD},n}$ from the model CMD, so the likelihood is calculated as follows:

\begin{align}
 \mathcal{L} = \prod_{n} \f{p_{\mathrm{M31},n} \times \mathcal{L}_{\mathrm{CMD},n}}{\eta(\textit{g}_{0,n}) \times \eta(\textit{i}_{0,n}) \times \eta(\textit{NB515}_{0,n})}.
\end{align}

Assuming that the prior distribution of the parameters is the uniform distribution of the intervals shown in Table \ref{table:priors}, the Python module \texttt{emcee} \citep{foreman-mackey2013} is used to estimate the distance for each subregion of each substructure. We initialize the sampler with 10 walkers, with 100000 steps after a burn-in period of 400000 steps to ensure a good sampling of the posterior distributions.

\begin{table}
 \caption{Priors of model parameters.}
 \label{table:priors}
 \begin{tabular*}{\columnwidth}{@{}l@{\hspace*{100pt}}l@{}}
       \hline
       parameters & intervals \\
       \hline
       $a$ & [0,2] \\
       $d$ [kpc] & [600,1100] \\
       $[\mathrm{Fe/H}]_0$ [dex] &  [-2.5,0] \\
       $\omega_{\mathrm{RGB}}$ [dex] &  [0,2] \\
       $b$ & [0,1]  \\
       \hline
 \end{tabular*}
\end{table}

\subsection{Metallicity Estimation}\label{subsection:MetallicityEstimation}

The photometric metallicities of individual stars are estimated by comparing the positions of stars on the CMD with the Dartmouth isochrones \citep{dotter2008} with given metallicities as follows. To estimate the photometric metallicity for each star, we construct 31 isochrones with $-2.5 < [\mathrm{Fe/H}] < +0.5$ in 0.1 dex intervals. To simplify the comparison with the previous photometric metallicity studies \citeg{ibata2014,mcconnachie2018}, we assume that the age is $13$ Gyr and alpha-abundance is $[\mathrm{\al/Fe}]=+0.0$ which is the same as the previous studies. It is noted that the alpha abundance of the M31 halo is not zero \citep[${\rm [\alpha/Fe]} > 0.0$][]{wojno2023}, so our metallicity estimation may underestimate by $< 0.2$ dex, systematically.

\begin{figure}
    \includegraphics[width=\columnwidth]
    {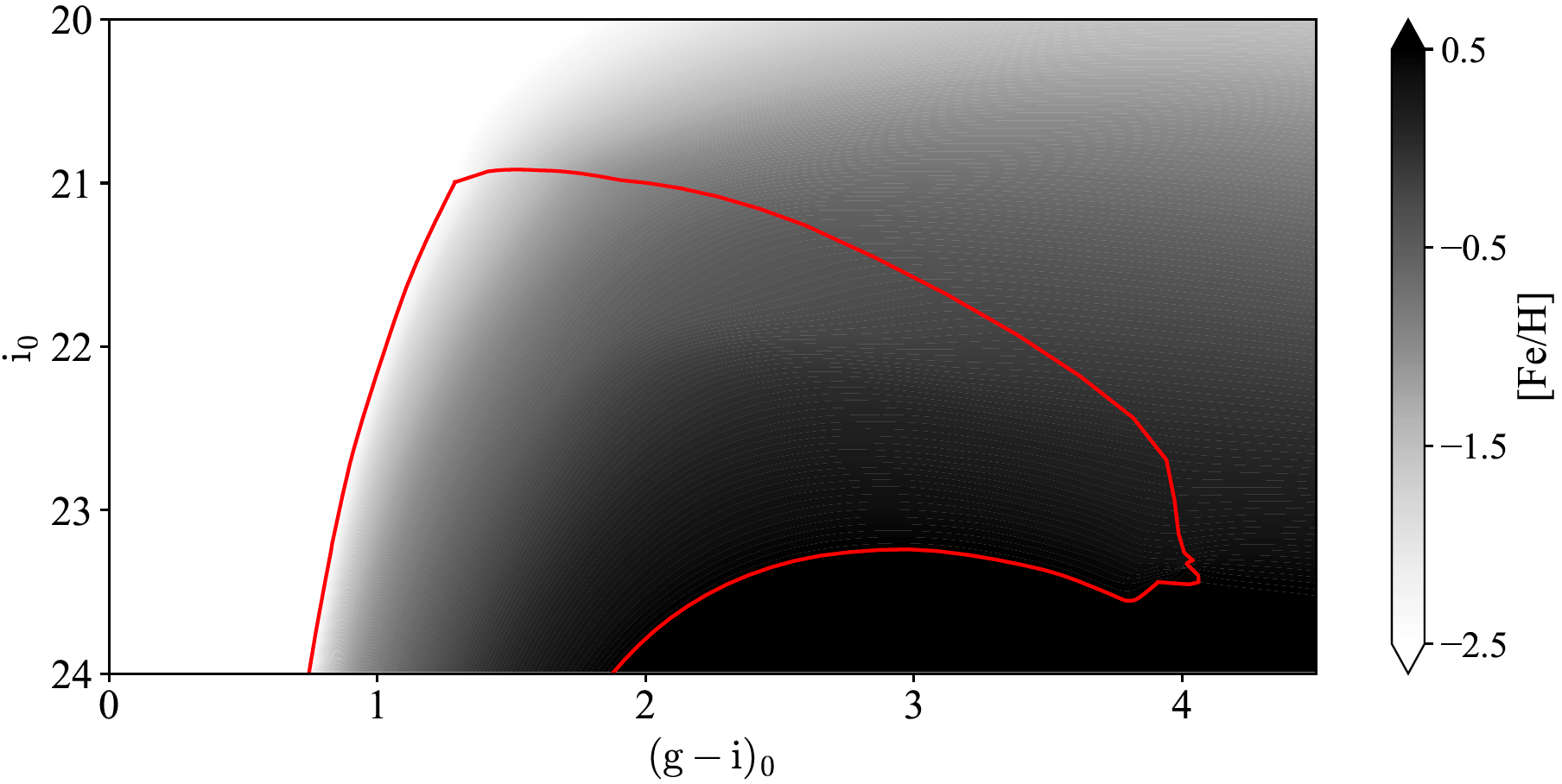}
    \caption{Interpolated metallicity map on the CMD for the M31 stellar halo. The red solid lines, which are the leftmost and rightmost, show the isochrones of the most metal-poor ($[\mathrm{Fe/H}]=-2.5$) and metal-rich isochrones ($[\mathrm{Fe/H}]=+0.50$). The line connecting the brightest ends of these lines corresponds to TRGB in the metallicity $-2.5<[\mathrm{Fe/H}]<+0.5$.}
    \label{figure:MDmodel}
\end{figure}

For the metallicity estimation, we also assume the distance to M31 to be 776 kpc \citep{dalcanton2012} and construct the metallicity model of RGB stars on the CMD by using the radial basis function (Rbf) interpolation of the \texttt{Python/SciPy} package. Figure \ref{figure:MDmodel} shows the interpolated RGB metallicity model. The metallicity of each star at a given position $((g-i)_0, i_0)$ on the CMD is estimated by interpolating [Fe/H] at that position in the metallicity model. By this method, we estimate the photometric metallicity of all point sources.

\section{Results}\label{section:results}
In this section, we describe the results of our analysis of the M31 halo, based on the catalogue constructed in the previous section. We describe the properties of the substructures in the M31 halo in Section \ref{subsection:substructures}. and the natures of the stellar halo in the entire analysis region in Section \ref{subsection:global}. 

\subsection{The properties of M31's substructures}\label{subsection:substructures}

\subsubsection{The spatial distribution for M31's substructures}\label{subsubsection:spatial}

\begin{figure*}
 \includegraphics[width=2\columnwidth]
 {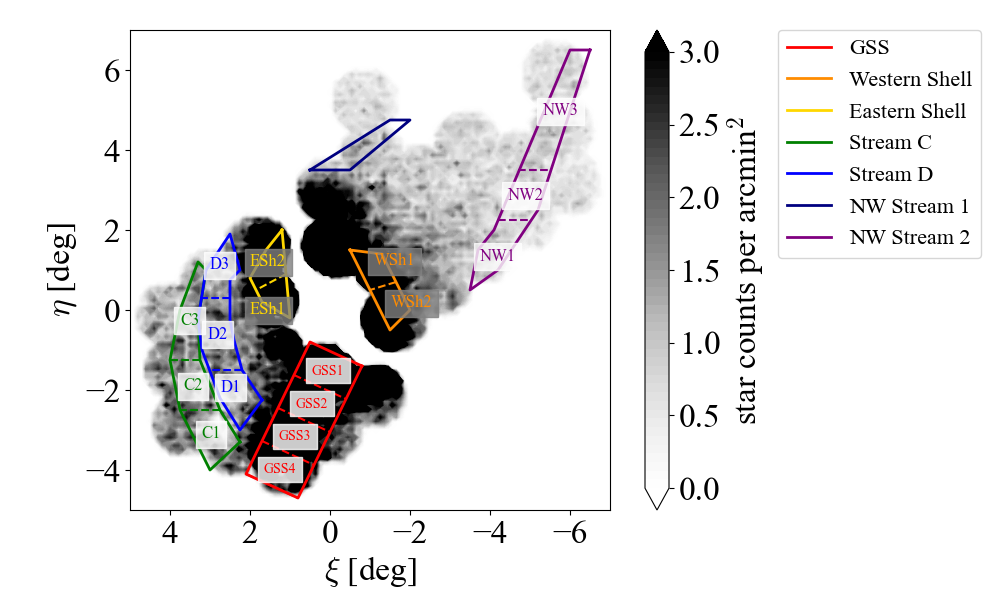}
 \caption{Stellar density map of the NRGB stars. The fields enclosed by colour-coded solid lines are the regions where we perform the distance estimation. The colour-coded dashed lines represent a subregion for distance estimation. In this work, we perform the distance estimation for each subregion.}
 \label{figure:SpatialDistribution}
\end{figure*}

\begin{figure*}
 \includegraphics[width=2\columnwidth]
 {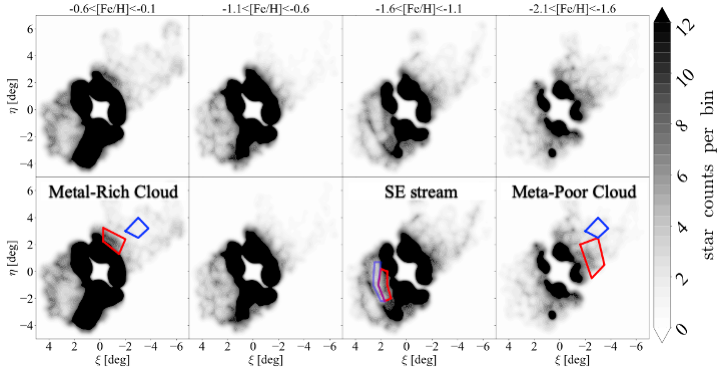}
 \caption{Stellar density map of NRGB stars with each range of the photometric metallicity. Each panel is constructed by smoothing with a Gaussian kernel with widths of 1.8 arcmin. The red boxes indicate the new substructures discovered in this study (Metal-Rich Cloud, SE Stream, and Metal-Poor Cloud from left (higher metallicity) to right (lower metallicity)). The blue boxes are assumed to be off-substructure fields to calculate the signal-to-noise ratio of newly discovered substructures.}
 \label{figure:RGB_Substructure}
\end{figure*}

The spatial distribution of the NRGB stars is shown in Figure \ref{figure:SpatialDistribution} and \ref{figure:RGB_Substructure}. Figure \ref{figure:SpatialDistribution} shows the spatial distribution of stars with high $p_{\mathrm{M31}}$ (i.e. including those red stars excluded in the NRGB colour-magnitude selection) and the substructures that have already been found. Fields of substructures are indicated by enclosing them with polygons which are determined visually. Figure \ref{figure:RGB_Substructure} shows the spatial distribution smoothed with a Gaussian kernel over 3 pixels ($= 1.8$ arcmin) and divided by [Fe/H] into four ranges (-0.6 to -0.1, -1.1 to -0.6, -1.6 to -1.1, and -2.1 to -1.6). In these maps, we can see the substructures (e.g., GSS, Eastern and Western Shell, Stream C and D, and NW Stream) that have already been found in previous studies. In Section \ref{subsubsection:distances}, we estimate the parameters (e.g., distance and metallicity) of the substructures that have already been found.

\begin{figure*}
 \includegraphics[width=2\columnwidth]
 {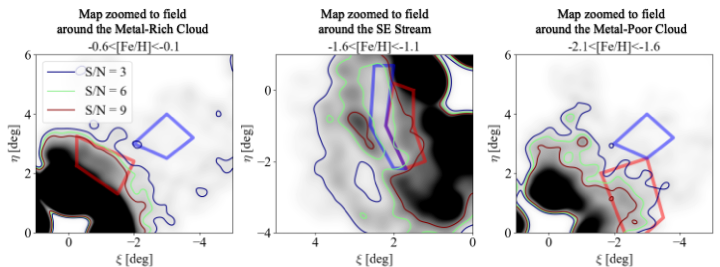}
 \caption{The same as in Figure \ref{figure:RGB_Substructure} but for the zoom-in stellar density map of the M31 halo with each photometric metallicity. The contours represent a signal-to-noise ratio (S/N) of 9 (dark red), 6 (light green), and 3 (navy).}
 \label{figure:RGB_Substructure_SN}
\end{figure*}

The substructures already found show a spatial distribution in good agreement with previous studies. In addition, we can identify the two overdense structures on the northwest halo at $(\xi, \eta) \sim (-1\degr, 2\fdg5)$ in the most metal-rich map and at $(\xi, \eta) \sim (-3\degr, 1\degr)$ in the most metal-poor map (hereafter, we call them the Metal-Rich Cloud and the Metal-Poor Cloud). Moreover, we can confirm one stream-like structure on the southeast halo at $(\xi, \eta) \sim (1\fdg5, -1\degr)$ (hereafter, we call this stream-like structure the SE Stream) from the visual inspection of the Gaussian kernel-smoothed maps (the red boxes in Figure \ref{figure:RGB_Substructure} indicates the newly discovered substructures). The blue boxes in Figure \ref{figure:RGB_Substructure} are assumed to be off-substructure fields to calculate the signal-to-noise ratio of newly discovered substructures. Figure \ref{figure:RGB_Substructure_SN} is a zoomed-in view of the newly discovered substructure field. The red and blue boxes are the same as Figure \ref{figure:RGB_Substructure}. The contours represent a signal-to-noise ratio (S/N) of 9 (dark red), 6 (light green), and 3 (navy), where S/N is estimated through a formula \texttt{$=$ (smoothed map - mean of off-substructure field) / standard deviation of off substructure field} \citep{yang2023}. As can be seen from this figure, the new substructures (red boxes) have a S/N of 6 or higher for the off-substructure fields (blue boxes), and these high S/N are the almost same as Stream C ($(\xi,\eta) \sim (2\fdg5,-1\degr)$ in the middle panel of Figure \ref{figure:RGB_Substructure_SN}) and NW Stream 1 ($(\xi,\eta) \sim (-0\fdg5,4\degr)$ in the right panel of Figure \ref{figure:RGB_Substructure_SN}). 

\begin{figure*}
 \includegraphics[width=2\columnwidth]
 {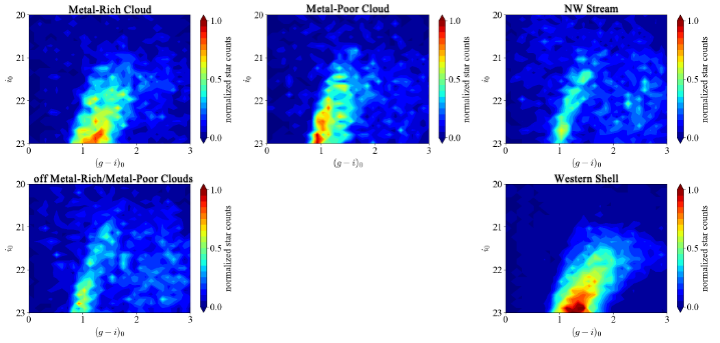}
 \caption{Colour-magnitude diagrams of the Metal-Rich, Metal-Poor Clouds, off Metal-Rich/Metal-Poor region, Western Shell and NW Stream.The objects used to construct these diagrams are M31-like objects with $p_{M31}>0.9$, and the number of objects in all diagrams is normalized.}
 \label{figure:MRC_MPC_CMD}
\end{figure*}

% About Metal-Rich/Metal-Poor Clouds
Figure \ref{figure:MRC_MPC_CMD} shows the Hess diagram with normalized number of stellar objects of Metal-Rich, Metal-Poor Clouds (fields defined by the red boxes in Figure \ref{figure:RGB_Substructure}), off Metal-Rich/Metal-Poor Clouds region (defined by the blue boxes in Figure \ref{figure:RGB_Substructure}), Western Shell and NW Stream. This figure shows the well-populated RGB sequence for Metal-Rich/Metal-Poor Clouds, in contrast to the off-clouds. The spatial distribution of both clouds slightly overlaps each other and it seems that most part of these new overdensities are an outward extension of the Western Shell. Also, for the Metal-Rich Cloud, NGC205-loop and North Spur are nearby, but this new one is located outside of these known structures. Only Andromeda XVI overlaps the Metal-Rich Cloud, but this dwarf galaxy is more metal-poor \citep[$\mathrm{[Fe/H]} \sim -2$;][]{letarte2009,collins2014,collins2015} than the Metal-Rich Cloud ($-0.6<[\mathrm{Fe/H}]<-0.1$). Therefore, although it is difficult to distinguish this cloud from Andromeda XVI using only spatial distribution, metallicity (i.e., the colour-magnitude information) can separate them. In addition to the Metal-Rich Cloud, the photometric metallicity distribution of the Western Shell seems to be wider than that of Metal-Poor Cloud in Figure \ref{figure:MRC_MPC_CMD}, so we can consider that this substructure is not part of the Western Shell, according to the metallicity information. Based on the above, we consider these substructures to be new discovered ones. In Section \ref{subsection:newsubstructures}, we describe distance estimation results for these clouds, and discuss the origin of these structures.

\begin{figure}
 \includegraphics[width=\columnwidth]
 {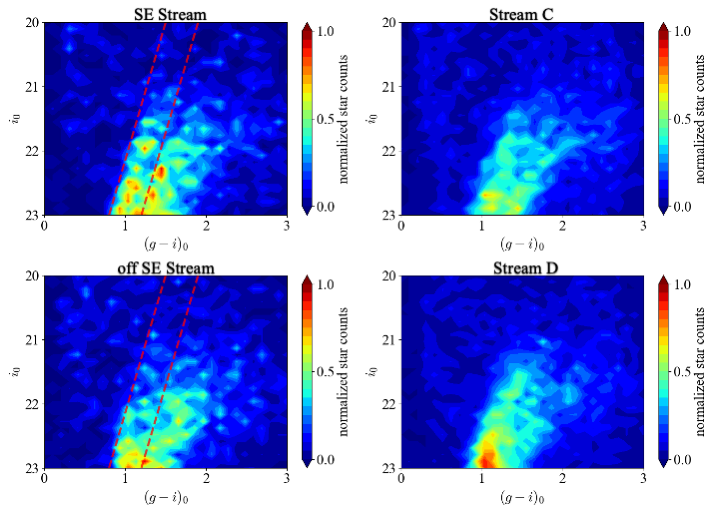}
 \caption{The same as in Figure \ref{figure:MRC_MPC_CMD} but for the CMDs of the SE Stream, off-SE Stream, Stream C, and Stream D. In the CMD of the SE Stream (upper left), RGB overdensity is seen within the red dashed lines, but no RGB overdensity is seen within this region in the CMD of off-SE Stream (lower left).}
 \label{figure:SEStream_CMD}
\end{figure}

% About SE Stream
Figure \ref{figure:SEStream_CMD} shows the Hess diagram with the normalized number of stellar objects of the SE Stream, off-SE Stream, Stream C, and Stream D. In this figure, the RGB sequence of the SE Stream extends to the red side, and metal-poor sequences such as Stream C and D also can be seen. In particular, the CMD of the off-Stream has no noticeable sequence within the area enclosed by the red dashed lines, while the SE Stream has a well-populated RGB sequence even within the red dashed lines. Therefore, it can be confirmed that there are stellar populations that are a combination of off-Stream components (i.e. field halo stars) and Stream C and/or D components (i.e. accreted stars; shown in the left panels of Figure \ref{figure:SEStream_CMD}). The South-Eastern Shell \citep[SE shell; ][]{gilbert2007} is very close to this stream. However, since this structure is slightly outside of the SE shell and is more metal-poor ($[\mathrm{Fe/H}]\sim-1.3$) than the SE shell ($[\mathrm{Fe/H}]\sim -0.50$), we can consider that the SE Stream is different from the SE shell. In Section \ref{subsection:newsubstructures}, we describe distance estimation results for these clouds, and discuss the origin of this structure, as well as Metal-Rich/Metal-Poor Clouds.

\subsubsection{Line-of-sight distances to M31's substructures}\label{subsubsection:distances}

We apply the method described in Section \ref{subsection:algorithm} to estimate the distances of the halo substructures of M31. Figure \ref{figure:SpatialDistribution} shows the density map of the NRGB stars. The solid colour-coded lines are the outer frame for each of the M31 substructure in the analysis region. The dashed colour-coded lines correspond to the subregions into which the substructures are divided for the distance estimation: to derive the distance and its gradient in each substructure, we perform distance estimation for each subregion. Note that an overlapping system of fields was implemented such that a given subregion XY (e.g., GSS12 shown in Table \ref{table:distance}) contains the stars from both subregion X and subregion Y. In addition, it should be noted that all stars with $0.5 < (\textit{g}-\textit{i})_0<2.5$ and $20<\textit{i}_0<22$ are used in this analysis when the distance estimation. Finally, we should note that the regions for substructures are determined visually, so they may deviate slightly from the regions estimated in previous studies.

\begin{table}
  \caption{The estimated distances of M31 substructures.}
  \label{table:distance}
  \begin{tabular*}{\columnwidth}{@{}l@{\hspace*{8pt}}l@{\hspace*{8pt}}l@{\hspace*{8pt}}l@{\hspace*{8pt}}l@{}}
      \hline
      substructure & subregion & distance [kpc] & [Fe/H] [dex] & $\sigma_{[\mathrm{Fe/H]}}$ [dex]\\
      \hline
      GSS & GSS1 & $769.14_{-3.54}^{+0.6}$ & $-0.43_{-0.02}^{+0.04}$ & $0.70_{-0.01}^{+0.01}$\\
       & GSS12 & $794.40_{-0.05}^{+3.62}$ & $-0.20_{-0.06}^{+0.07}$ & $0.88_{-0.03}^{+0.03}$\\
       & GSS2 & $812.87_{-0.04}^{+0.06}$ & $-0.06_{-0.13}^{+0.18}$ & $0.96_{-0.06}^{+0.06}$\\
       & GSS23 & $812.89_{-0.06}^{+0.10}$ & $-0.21_{-0.12}^{+0.14}$ & $0.92_{-0.06}^{+0.05}$\\
       & GSS3 & $805.47_{-0.07}^{+7.40}$ & $-0.09_{-0.11}^{+0.13}$ & $0.95_{-0.04}^{+0.05}$\\
       & GSS34 & $835.79_{-0.13}^{+0.32}$ & $-0.50_{-0.10}^{+0.12}$ & $0.81_{-0.05}^{+0.06}$\\
       & GSS4 & $835.76_{-0.12}^{+0.22}$ & $-0.09_{-0.21}^{+0.28}$ & $0.95_{-0.09}^{+0.11}$\\
      \hline
      Eastern Shell & ESh1 & $774.79_{-2.09}^{+1.48}$ & $-0.41_{-0.04}^{+0.05}$ & $0.64_{-0.02}^{+0.02}$\\
       & ESh12 & $770.21_{-1.04}^{+0.75}$ & $-0.39_{-0.03}^{+0.03}$ & $0.66_{-0.02}^{+0.02}$\\
       & ESh2 & $765.62_{-0.01}^{+0.03}$ & $-0.38_{-0.04}^{+0.04}$ & $0.69_{-0.02}^{+0.02}$\\
      \hline
      Western Shell & WSh1 & $765.66_{-0.06}^{+0.08}$ & $-0.49_{-0.06}^{+0.07}$ & $0.76_{-0.03}^{+0.03}$\\
       & WSh12 & $772.87_{-0.13}^{+0.21}$ & $-0.40_{-0.10}^{+0.10}$ & $0.79_{-0.05}^{+0.05}$\\
       & WSh2 & $783.54_{-0.10}^{+0.14}$ & $-0.45_{-0.06}^{+0.07}$ & $0.78_{-0.03}^{+0.03}$\\
      \hline
      NW Stream & NW1 & $728.72_{-13.08}^{+6.77}$ & $-1.30_{-0.04}^{+0.05}$ & $0.37_{-0.03}^{+0.03}$\\
       & NW12 & $743.94_{-6.94}^{+0.64}$ & $-1.35_{-0.03}^{+0.04}$ & $0.38_{-0.02}^{+0.02}$\\
       & NW2 & $744.99_{-19.26}^{+7.38}$ & $-1.41_{-0.05}^{+0.05}$ & $0.38_{-0.03}^{+0.03}$\\
       & NW23 & $742.89_{-7.81}^{+70.68}$ & $-1.38_{-0.09}^{+0.05}$ & $0.34_{-0.04}^{+0.03}$\\
       & NW3 & $809.13_{-67.50}^{+19.84}$ & $-1.41_{-0.07}^{+0.09}$ & $0.31_{-0.03}^{+0.04}$\\
      \hline
      Stream C & C1 & $769.28_{-44.70}^{+7.41}$ & $-0.85_{-0.06}^{+0.15}$ & $0.57_{-0.04}^{+0.05}$\\
       & C12 & $773.54_{-10.46}^{+6.76}$ & $-0.89_{-0.05}^{+0.07}$ & $0.58_{-0.03}^{+0.03}$\\
       & C2 & $820.93_{-44.06}^{+10.91}$ & $-0.96_{-0.07}^{+0.15}$ & $0.64_{-0.03}^{+0.07}$\\
       & C23 & $798.01_{-21.48}^{+23.25}$ & $-0.83_{-0.12}^{+0.12}$ & $0.68_{-0.04}^{+0.06}$\\
       & C3 & $820.69_{-3.30}^{+0.67}$ & $-1.06_{-0.03}^{+0.04}$ & $0.54_{-0.02}^{+0.03}$\\
      \hline
      Stream D & D1 & $769.33_{-3.08}^{+0.05}$ & $-1.18_{-0.02}^{+0.02}$ & $0.49_{-0.02}^{+0.02}$\\
       & D12 & $769.72_{-0.53}^{+2.99}$ & $-1.17_{-0.02}^{+0.02}$ & $0.50_{-0.02}^{+0.02}$\\
       & D2 & $772.94_{-0.24}^{+18.15}$ & $-1.02_{-0.07}^{+0.04}$ & $0.56_{-0.02}^{+0.03}$\\
       & D23 & $809.70_{-3.79}^{+7.93}$ & $-1.00_{-0.04}^{+0.04}$ & $0.57_{-0.02}^{+0.02}$\\
       & D3 & $817.20_{-7.94}^{+10.85}$ & $-1.04_{-0.03}^{+0.03}$ & $0.53_{-0.01}^{+0.02}$\\
      \hline
      \end{tabular*}
\end{table}

Table \ref{table:distance} is the results of the estimation for each of the M31 substructures, showing the distance, metallicity, and metallicity deviation, and their random errors (corresponds to 68 $\%$ Bayesian credible interval for posterior distribution). In addition to random error, we estimate the systematic error of the distance as follows. As described in Section \ref{subsection:algorithm}, the distance estimation is performed by model fitting on the CMD, in which there exist several systematic errors. First, according to \citet{tonry2012}, Pan-STARRS 1 has a systematic error of 0.02 mag ($\sim 7$ kpc) for each \textit{g}- and \textit{i}-bands. PAndAS data used the PanSTARRS photometry to correct for zero-point uncertainties in the survey region, so these systematic photometric errors are included as total systematic errors of distance estimation. Second, as described in Section \ref{subsection:data}, we correct for the magnitude based on the dust map by Schlegel et al., and extinction correction is considered to have a systematic error of about $10~\%$, which corresponds to a magnitude of 0.015 mag ($\sim$ 5 kpc). Third, in this distance estimation method using the Dartmouth isochrones, we find that the distance changes by a maximum of 5 kpc, if we change the age assumption to 2 Gyr ($10 \pm 2$ Gyr). Finally, \citet{serenelli2017} showed that RGB stars in such a theoretical curve contain a systematic error of 0.05 mag ($\sim 10 $kpc). Adding these four factors, we estimate the total systematic error in the estimated distance to be $\sim 27$ kpc. 

% GSS
\begin{figure*}
\includegraphics[width=2\columnwidth]
{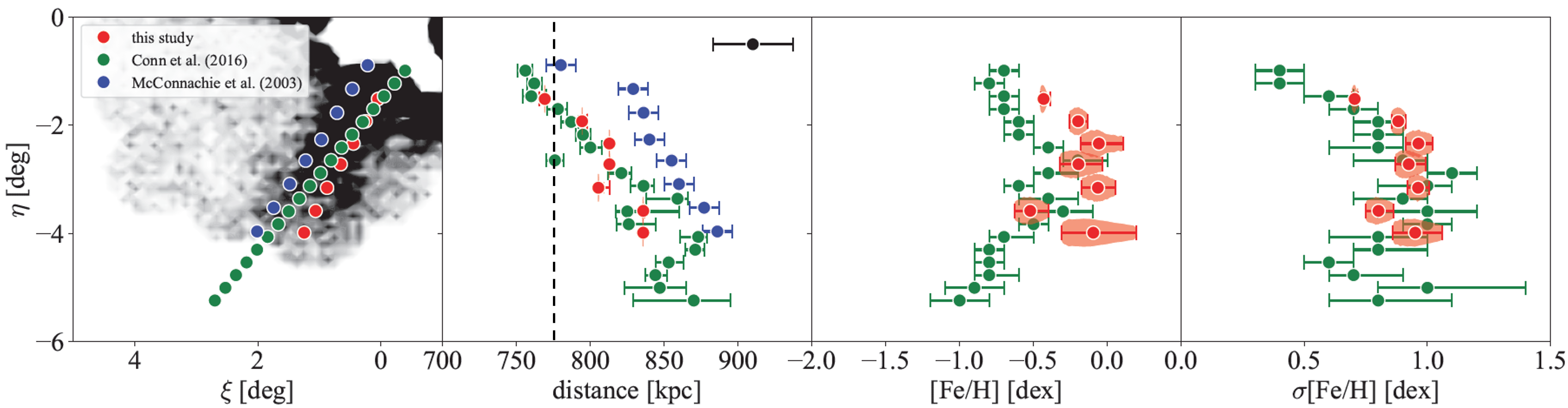} 
\caption{Results of the distance estimation for the GSS. Leftmost: the spatial distribution with points where we conduct distance estimation. Inner-left: the distance distribution as a function of $\eta$. Inner-right: the photometric metallicity of the stellar systems. The vertical black dashed line shows the line-of-sight distance to the M31 main body \citep[776 kpc][]{dalcanton2012} Rightmost: the standard deviation of the metallicity. In each panel, red points as violin plots show the results of this study, and green and blue dots correspond to the results of \citet{conn2016} and \citet{mcconnachie2003}, respectively.}\label{figure:distance_GSS}
\end{figure*}

Figure \ref{figure:distance_GSS} shows the results of distance estimation for GSS. The leftmost panel is the spatial distribution with points where these are centers of stars within the distance-estimated region weighted by $p_{M31}$. The inner-left panel is the result of the distance as a function of $\eta$ coordinate. The inner-right is the estimated metallicity of stellar systems, and the rightmost panel shows the standard deviation of the metallicity. In each panel, red dots show the results of this study, and green \& blue dots correspond to the results of \citet{conn2016} and \citet{mcconnachie2003}, respectively. In the three figures to the right, the results of this study as red points are displayed by violin plots, where the shaded region represents the shape of the posterior distribution within the $68\%$ Bayesian credible interval. A black dot shows systematic error in the estimation of this study. The results for GSS are generally consistent with previous studies, and we can confirm the same distance gradient that is reported by previous studies. In addition, the results of estimated metallicity and its standard deviation are also almost consistent with those of \citet{conn2016}. 
While the overall trend is generally consistent, some subregions (e.g., GSS2, GSS3, GSS4) are estimated to be more metal-rich than those of \citet{conn2016}. This discrepancy may be attributed to the limited number of objects used in the analysis. However, when using objects from both subregions, such as GSS23 and GSS34, which increases the sample size, the results show a good agreement with those of \citet{conn2016}. While generally consistent with the metal amounts estimated in previous works, when compared to recent spectroscopic results \cite[e.g.,][]{escala2021} shows that the metallicity distribution of GSS is more metal-poor ($[{\rm Fe/H}] < -0.8$) than our result as one moves away from the center of M31. This result may be due to the lack of correction for the kinematically hot stellar halo of M31 in the analysis of photometric data. Therefore, it is necessary to confirm these metallicity peaks using future follow-up spectroscopic observations as Subaru/Prime-Focus Spectrograph \citep[PFS;][]{takada2014}.

% Stream C and D
\begin{figure*}
\includegraphics[width=2\columnwidth]
{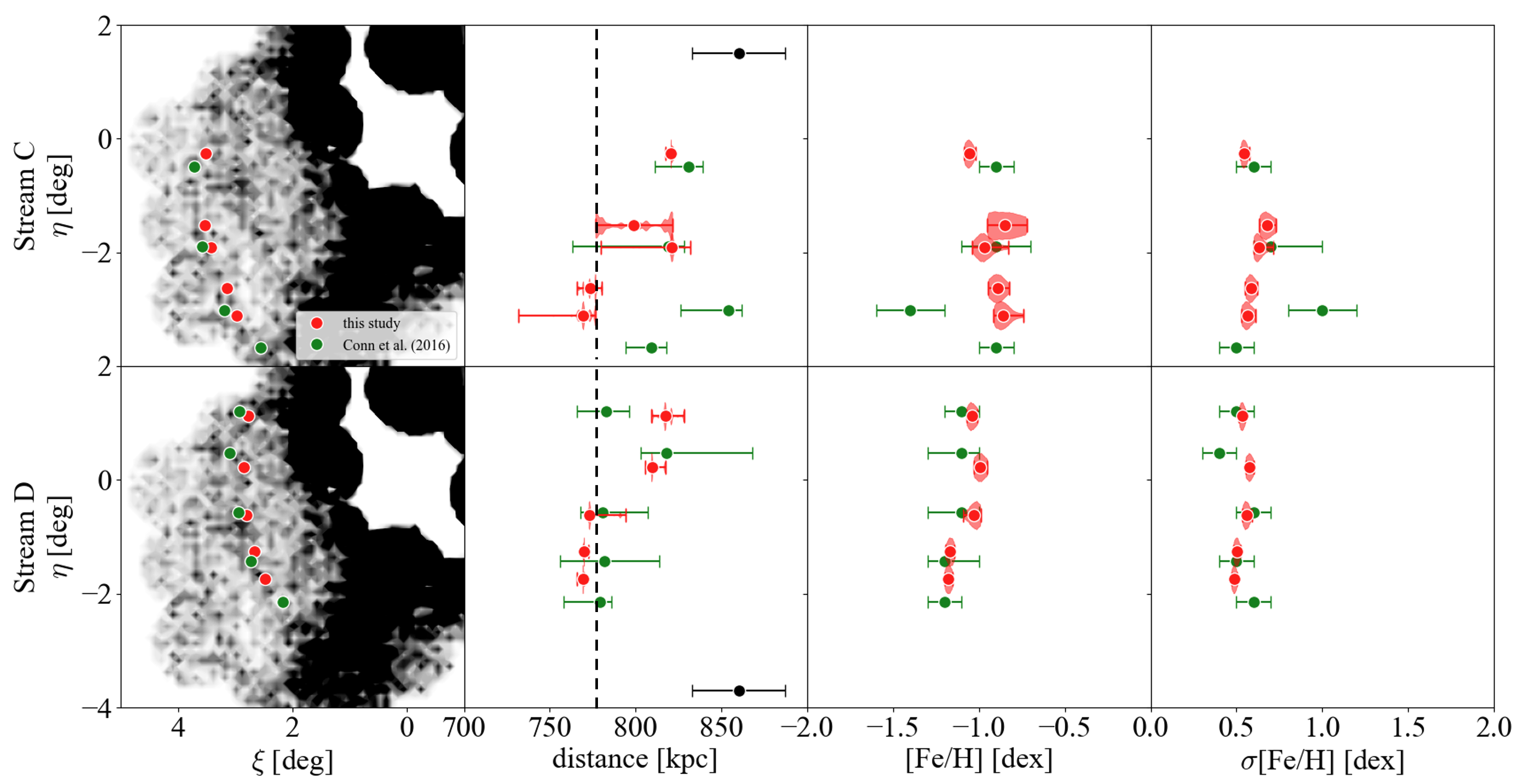} 
\caption{The same as in Figure \ref{figure:distance_GSS} but for the results of the distance estimation for the Stream C (top column) and the Stream D (bottom column). In each panel, red points show the results of this study, and green dots correspond to \citet{conn2016}.}\label{figure:distance_C_D}
\end{figure*}

Figure \ref{figure:distance_C_D} shows the results of the distance estimation for the Stream C and the Stream D, similar to Figure \ref{figure:distance_GSS}. The red points in each figure show the estimation results and errors of this study, and the green dots indicate the results of \citet{conn2016}. Although our results are basically in agreement with previous results, we identify the presence of significant distance gradients for both streams clearly. We think that these trends 
can be more clearly seen thanks to the addition of the narrow-band data in our selection of candidate member stars. In \citet{conn2016}, stars other than the M31 substructure are treated statistically based on the method of \citet{martin2013}, in which the contamination model was constructed based on regions where the M31 substructure was undetected. In contrast, we assign each star to be a contamination source or not, based on its \textit{NB515} magnitude (see Section \ref{subsection:selection_RGB}). Therefore, the difference in treating contamination sources is likely to be reflected in the distance estimation results and their precisions.

% NW Stream
\begin{figure*}
\includegraphics[width=2\columnwidth]
{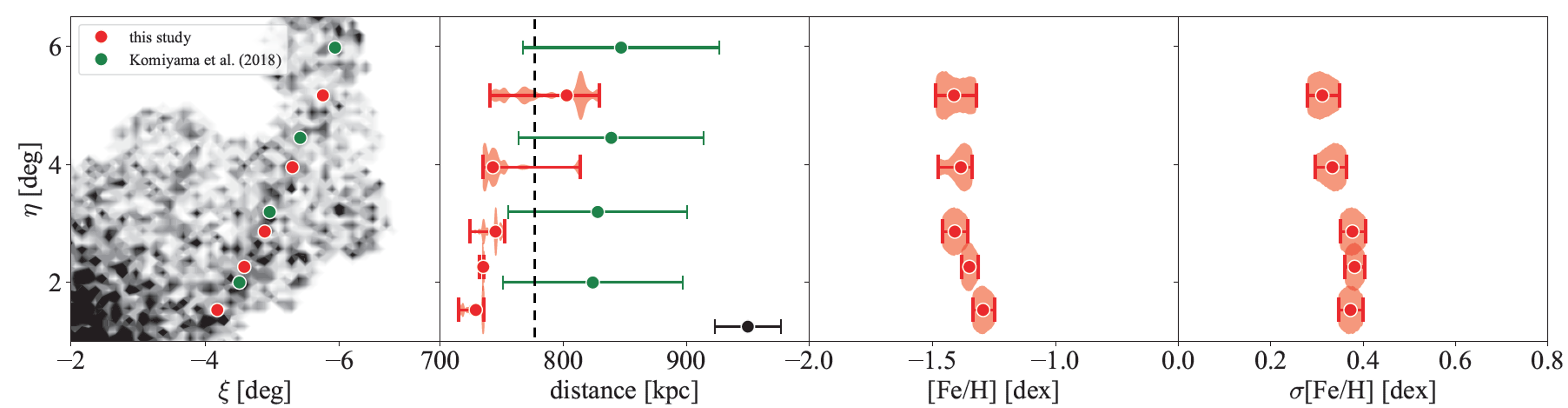} 
\caption{The same as in Figure \ref{figure:distance_GSS} but for the results of the distance estimation for the NW Stream. In each panel, red points show the results of this study, and green dots correspond to \citet{komiyama2018a}.}\label{figure:distance_NW}
\end{figure*}

Figure \ref{figure:distance_NW} shows the distance estimation results for the NW Stream. The red points and green dots are the results of this study and \citet{komiyama2018a}, respectively. In the region at $\eta>3$, the results agree with those of \citet{komiyama2018a} within the error range. However, \citet{komiyama2018a} concluded that the NW Stream is fully located behind M31 from their distance estimation using Red Clump stars (RC). On the other hand, our analysis suggests that the NW Stream is distributed in front of M31 at $\eta<3$ (where the distance is below 776 kpc). We think this difference occurs due to the choice of different distance indicators and whether contamination removal is performed. \citet{komiyama2018a} performed the estimation using RC, which differs from this study with RGB stars. In addition, they did not use \textit{NB515}, because the \textit{NB515} data is shallower than the magnitude of RC in the M31 halo. RC ($\delta i_0 \sim \pm0.05$) used in \citet{komiyama2018a} is a larger intrinsic ($\sim$ distance) error than that of RGB stars ($\delta i_0 \sim \pm 0.015$ mag), and they do not take into account the contamination sources (e.g., foreground stars and background unresolved galaxies). Considering that there is $\sim$ 20 kpc of systematic errors in addition to the random errors, the total errors in distance estimation are large and it is not possible to rule out the possibility that the stream is in front of M31. Also, the RC method by \citet{komiyama2018a} contains the effect of contamination sources, despite the fact that foreground stars are severely contaminated in the region where the NW Stream is located. Therefore, even the results of Komiyama et al. indicate that the NW Stream can be distributed in front of M31, taking both random and systematic errors into account.

In this study, the mean metallicity of NW Stream subregions is $[\mathrm{Fe/H}]=-1.4$. \citet{komiyama2018a} found that the RGB and RC stars in the CMD of the NW Stream could be traced by an isochrone with $[\mathrm{Fe/H}] = -1.37,~~~log{\mathrm{age}} = 10.00$), so the result of this study is consistent with that of previous study. Therefore, we can conclude the NW Stream is metal-poor.

% Eastern Shell and Western Shell
\begin{figure*}
\includegraphics[width=2\columnwidth]
{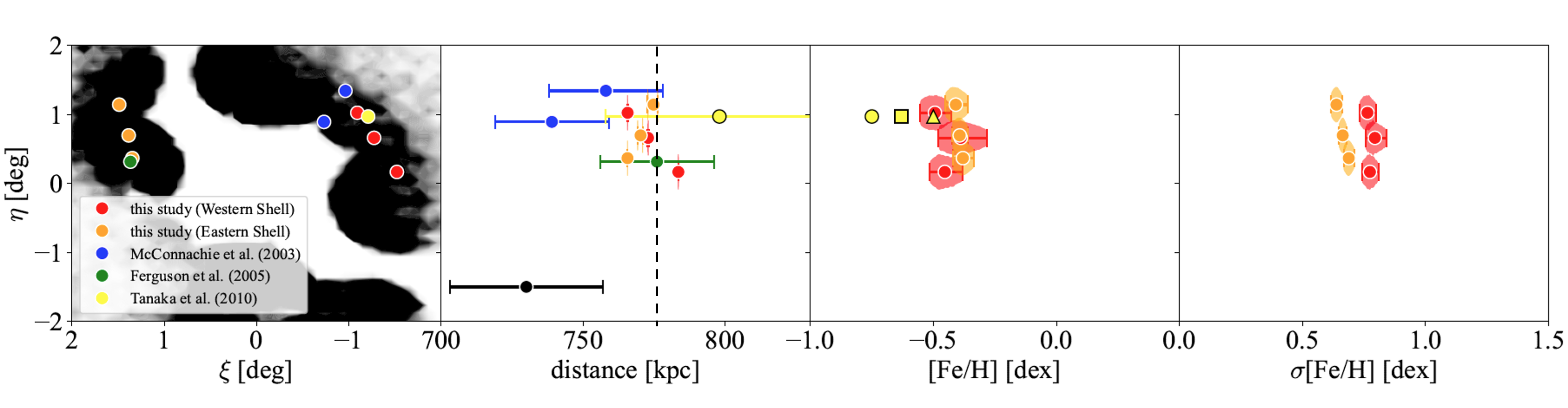} 
\caption{The same as in Figure \ref{figure:distance_GSS} but for the results of the distance estimation for the Western Shell and the Eastern Shell. In each panel, red and orange points show the results of Western Shell and Eastern Shell in this study, and blue, green, and yellow dots correspond to \citet{mcconnachie2003}, \citet{ferguson2005}, and \citet{tanaka2010}, respectively. In the figure in the middle right, the results of three different analyses of \citet{tanaka2010} are shown; the mean (filled circle), median (triangle), and peak (square) of the metallicity distribution of the Western Shell.}
\label{figure:distance_shell}
\end{figure*}

Figure \ref{figure:distance_shell} shows the results for the Eastern Shell and the Western Shell. The orange and red points correspond to the Eastern Shell and the Western Shell in this study. The blue, green, and yellow dots correspond to the previous studies \citep[][respectively]{mcconnachie2003,ferguson2005,tanaka2010}. For both shell structures, the distances are consistent with the previous studies within the error range. Also, the metallicities of these substructures are consistent with those of the GSS. In addition to being consistent with previous studies, the results of the Western Shell and the Eastern Shell are in good agreement with each other. Simulation studies have suggested that these two shell structures were formed during the accretion of the GSS progenitor \citep[e.g.,][]{fardal2006,fardal2007}. Other than these studies, this association is supported by the recent chemodynamical studies \citep{escala2022a,dey2023}. The estimated metallicities of the two shells in this study are the almost same value, and are also consistent with those of GSS. Therefore, these results emphasize that these two shells have the same origin as GSS.

% for all substructures
In the southeastern part of the M31 halo, the substructures are crowded (e.g., GSS, Stream C, Stream D, Eastern Extent), so it is difficult to separate them based on photometric data only. However, there are no differences between parts of the same structure. Therefore, we think that we focus on the same substructure between adjacent subregions without the effect of the other structures, when we see within certain substructure.

\subsection{The Global Halo Properties}\label{subsection:global}
In this section, we use the NRGB stars defined in Section \ref{subsection:selection_RGB} to show the metallicity distribution in Section \ref{subsubsection:MD} with its estimation method, and the surface brightness profile in Section \ref{subsubsection:SBprofile}. Note that NRGB stars are cut by the colour-magnitude information in addition to $p_{\mathrm{M31}} > 0.9$. In particular, it should be noted that the results shown in Section \ref{subsubsection:MD} are biased against metal-rich stars because stars with $(\textit{g}-\textit{i})_0<2.5$ are selected.

In some previous studies \citeg{gilbert2012,ibata2014}, metallicity distribution and surface brightness profiles were constructed by masking the substructures. However, it is difficult to mask the substructure with only the data used in this study. Therefore, it should be noted that the global halo profiles include substructures.

\subsubsection{Metallicity distribution}\label{subsubsection:MD}

\begin{figure}
    \includegraphics[width=\columnwidth]
    {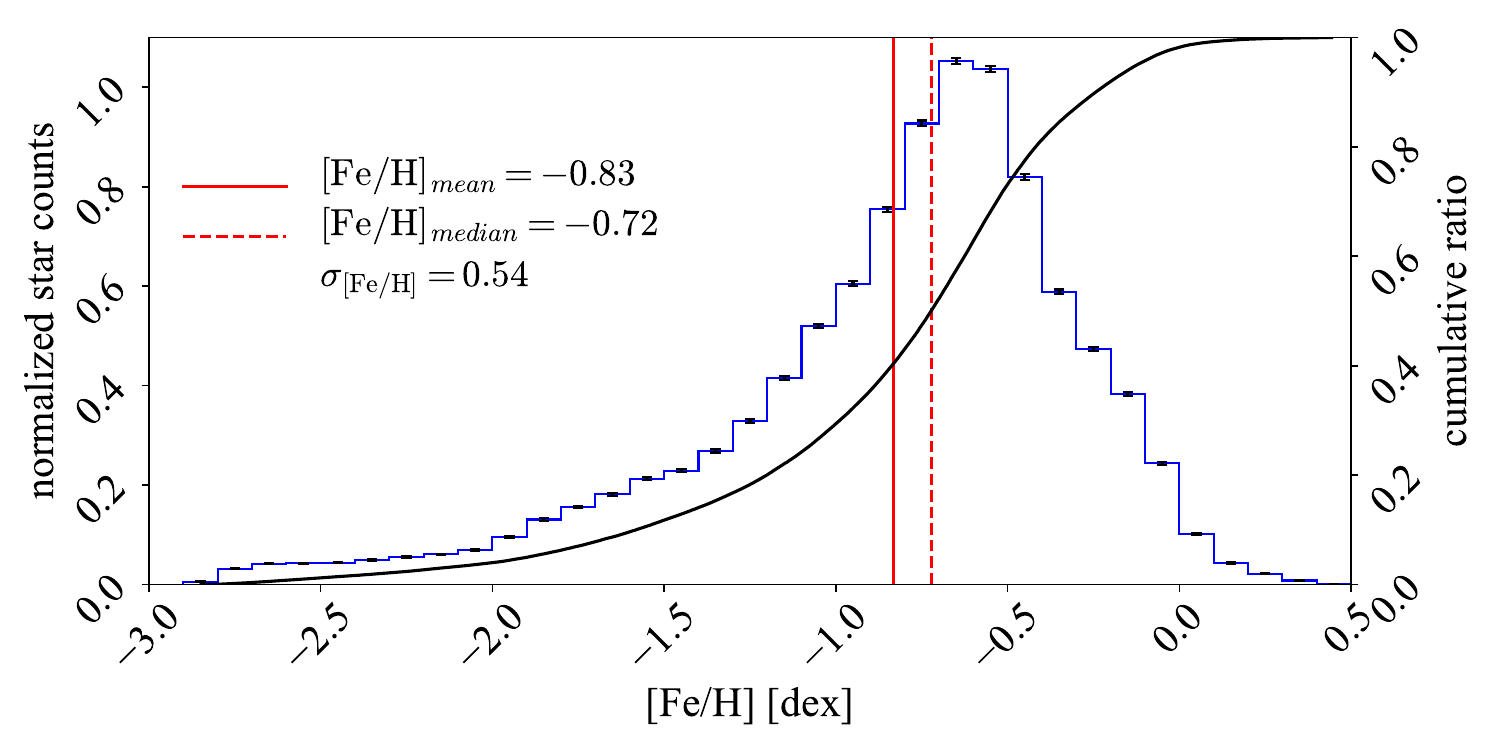}
    \caption{Metallicity distribution (blue histogram) of star with NRGB stars over the entire analysis region. The red solid (dashed) line shows the mean (median) value, and the black solid line indicates cumulative distribution. The black bars at each bin reflect the Poisson error. The left vertical axis shows the normalized star counts, and the right vertical axis shows the cumulative ratio.}
    \label{figure:MD_overall}
\end{figure}

We apply the method described in Section \ref{subsection:MetallicityEstimation} to estimate the photometric metallicity of the individual stars. As described in Section \ref{subsection:selection_RGB}, there are remaining dwarf stars with high $p_{\mathrm{M31}}$. Assuming the distance of M31, these stars are estimated to be metal-rich stars on the CMD. Therefore, we use NRGB stars to investigate the metallicity distribution of the M31 stellar halo without uncertainties due to foreground contaminations.

The estimated metallicity distribution of NRGB stars in the entire region is shown in Figure \ref{figure:MD_overall}. In Figure \ref{figure:MD_overall}, the black line is the cumulative distribution, the solid red line is the mean metallicity, and the dashed red line is the median value. The mean, median, and standard deviation of the metallicity for the entire analysis area are estimated to be $[\mathrm{Fe/H}]_{\mathrm{mean}} = -0.83$, $[\mathrm{Fe/H}]_{\mathrm{median}} = -0.72$, and $\sigma_{[\mathrm{Fe/H}]} = +0.54$, respectively. In this study, we truncate red stars with $(\textit{g}-\textit{i})_0>2.5$, when selecting the RGB-box (see Section \ref{subsection:selection_RGB}) to create the metallicity distribution.  Therefore, it should be noted that we may have underestimated the metal-rich side of our metallicity distribution.

\begin{figure*}
    \begin{center}
    \includegraphics[width=2\columnwidth]{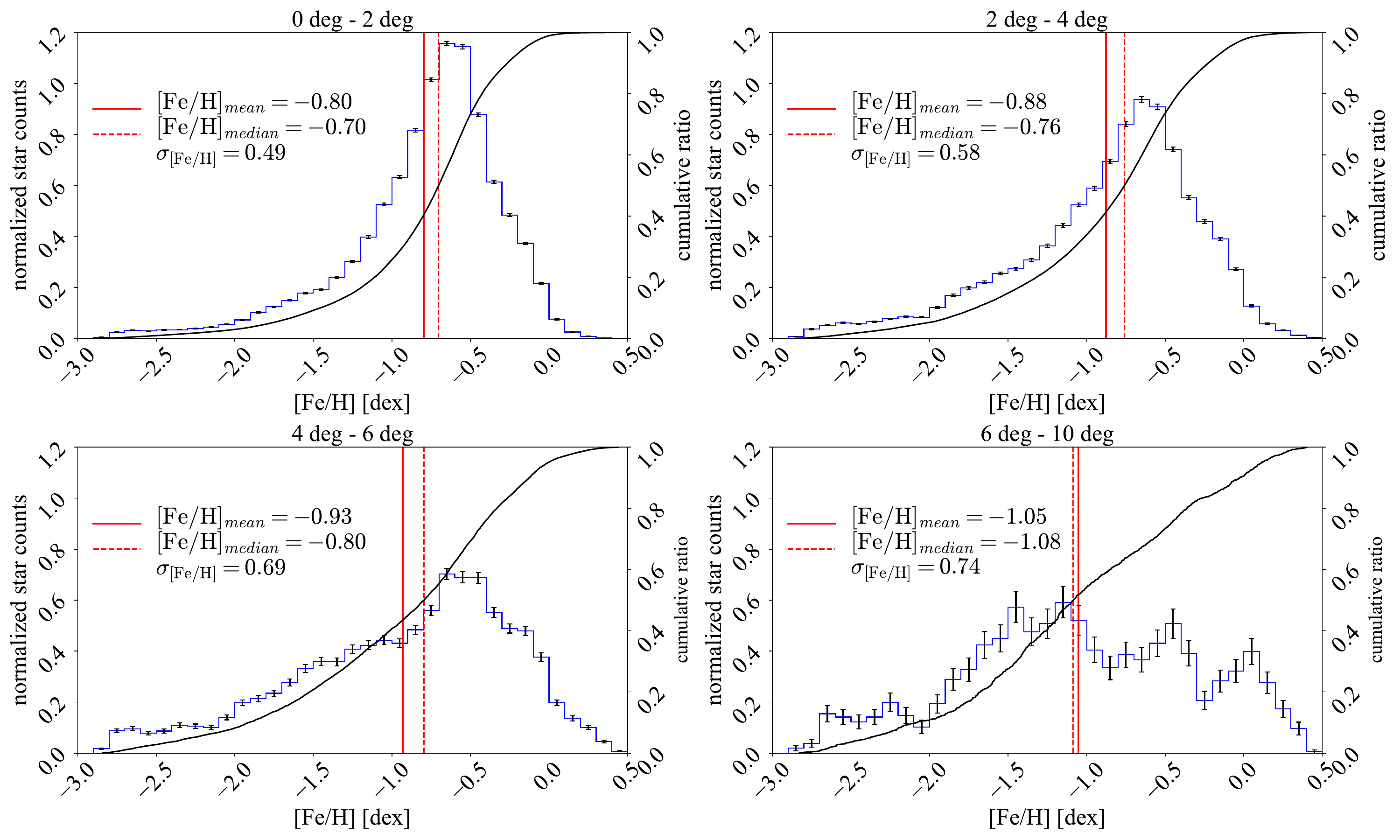}
    \end{center}
    \caption{Metallicity distribution in 4 annuli (Upper-left: 0 - 2 deg. Upper-right:  2 - 4 deg. Lower-left: 4 - 6 deg. Lower-right: 6 - 10 deg). These figures are almost the same as Figure \ref{figure:MD_overall}.}
    \label{figure:MD_annulus}
\end{figure*}

\begin{table}
 \caption{The photometric metallicity in four annuli.}
 \label{table:MD}
 \begin{tabular*}{\columnwidth}{@{}l@{\hspace*{30pt}}l@{\hspace*{30pt}}l@{\hspace*{30pt}}l@{}}
      \hline
      Projected radius & $[\mathrm{Fe/H}]_{\mathrm{mean}}$ & $[\mathrm{Fe/H}]_{\mathrm{median}}$ & $\sigma_{[\mathrm{Fe/H}]}$ \\
      deg & dex & dex & dex \\
      \hline
      $\mathrm{r_{proj}} = 0 - 2$ & -0.80 & -0.70 & 0.49\\
      $\mathrm{r_{proj}} = 2 - 4$ & -0.88 & -0.76 & 0.58\\
      $\mathrm{r_{proj}} = 4 - 6$ & -0.93 & -0.80 & 0.69\\
      $\mathrm{r_{proj}} = 6 - 10$ & -1.05 & -1.08 & 0.74\\
      \hline
 \end{tabular*}
\end{table}

Figure \ref{figure:MD_annulus} shows the metallicity distribution of NRGB stars in 4 annuli (projected radius $\mathrm{r_{proj}} = 0 - 2$ deg, $2 - 4$ deg, $4 - 6$ deg, and $6 - 10$ deg). The photometric metallicity results for each annulus are summarized in Table \ref{table:MD}. Table \ref{table:MD} shows that the outer annulus is more metal-poor, confirming the spatial metallicity gradient in the M31 stellar halo \citep{kalirai2006}. In the observed fields, the outer region is closer to the Galactic plane, which is more susceptible to contamination of the Galactic disc stars because it is known that the number of such stars increases with an exponential function \citep{tanaka2010,martin2013}. Galactic disc stars are generally identified as red stars (metal-rich populations assuming a distance of M31), so the metallicity distribution and metallicity profile of the M31 halo would have been estimated as metal-rich, especially in the northwestern region, unless the contamination sources are correctly removed. However, as shown above, the metallicity distribution in this study shows that the outer region of the M31 halo is more metal-poor than the inner region. This result indicates that \textit{NB515} is a powerful tool to eliminate foreground stars and that we are able to capture the true structure of the M31 stellar halo.

To further obtain the global metallicity property of the M31 stellar halo, we construct both the radial metallicity profile and the profiles along the southeast and northwest directions of the minor axis of the M31 disc (hereafter, we call these two profiles SE/NW minor axis profiles). In general, the minor axis profiles allow us to obtain the stellar halo profile with minimal influence of both the disc component and the stellar populations in a particular direction than the radial profile, because our analysis region is anisotropic. We construct three profiles using the NRGB stars as follows. First, the M31 halo is divided into small regions. In the case of the radial profile, the stars are divided into 3$\arcmin$ areas from the M31 centre throughout the analysis region: in the case of the minor axis profiles, the SE minor axis profile is constructed by the data in the southeastern region (PFS\_FIELD\_19 - PFS\_FIELD\_33; see Table \ref{table:obs_condition}) and the NW minor axis profile is from the data northwestern region (PFS\_FIELD\_1 - PFS\_FIELD\_16 and M31\_003, 004, 009, 022, 023; see Table \ref{table:obs_condition}). Using these data, the observed fields are divided into $3\arcmin \times 1\degr$ areas, to construct the minor axis metallicity profiles. Finally, we calculate the mean/median metallicity of the sources in each divided region. Then, the error of the mean/median metallicity in each region is the standard deviation of the metallicity of the objects. It should be the variation in the stellar metallicity within a region divided by the number of stars, but this value is sufficiently small because there are a sufficient number of sources within each region.

\begin{figure}
 \includegraphics[width=\columnwidth]
 {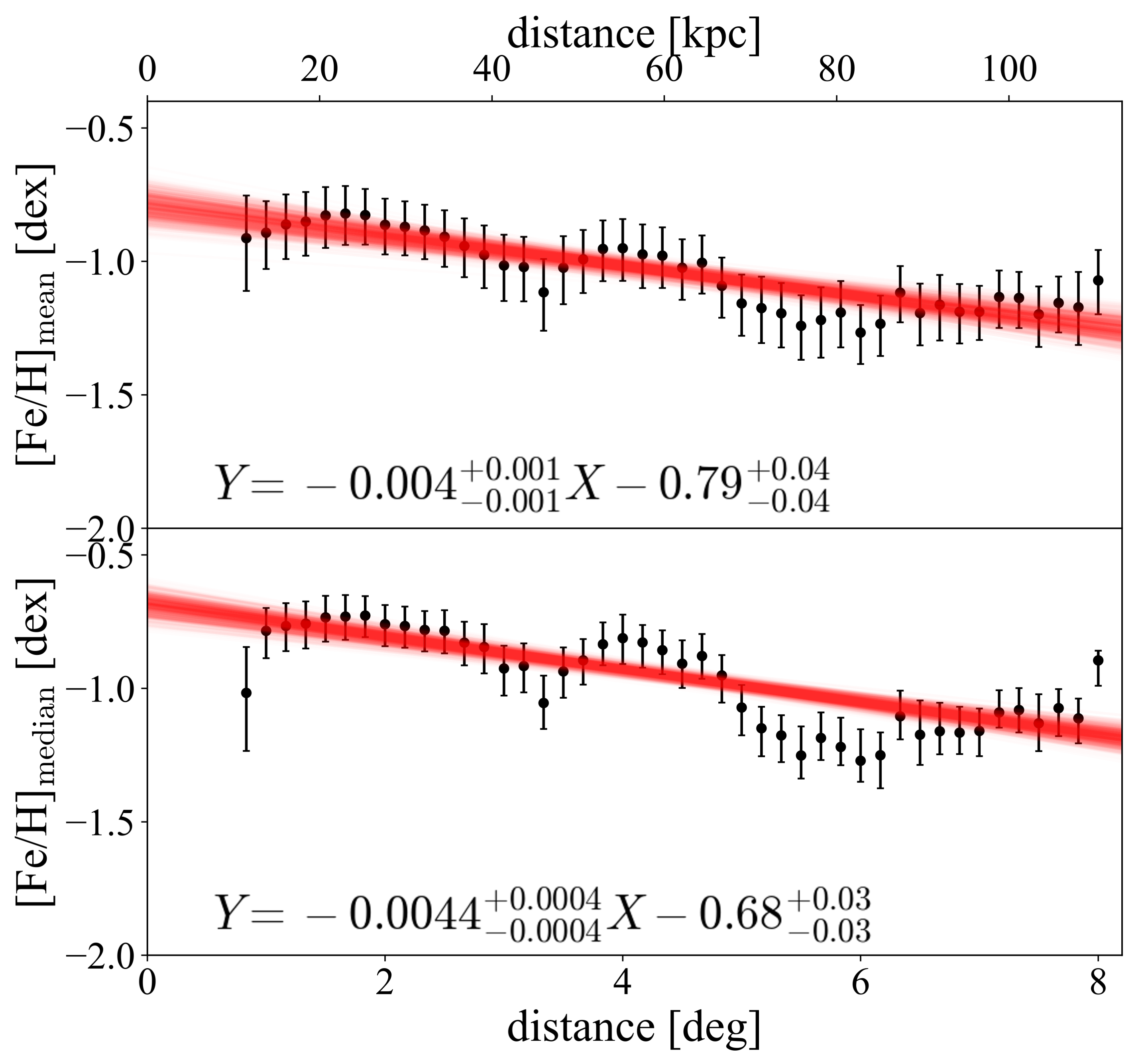}
 \caption{The metallicity profile along the radial direction from the M31 centre. The black dots in the top/bottom figures indicate the mean/median values of the metallicity in each region, and the red lines are 1000 linear regression lines randomly sampled from the posterior distribution. The horizontal axis represents the distance from the M31 centre, with the horizontal axis displayed below in units of [deg] and that displayed above in units of [kpc].}
 \label{figure:MD_RadialProfile}
\end{figure}

\begin{figure*}
 \includegraphics[width=2\columnwidth]
 {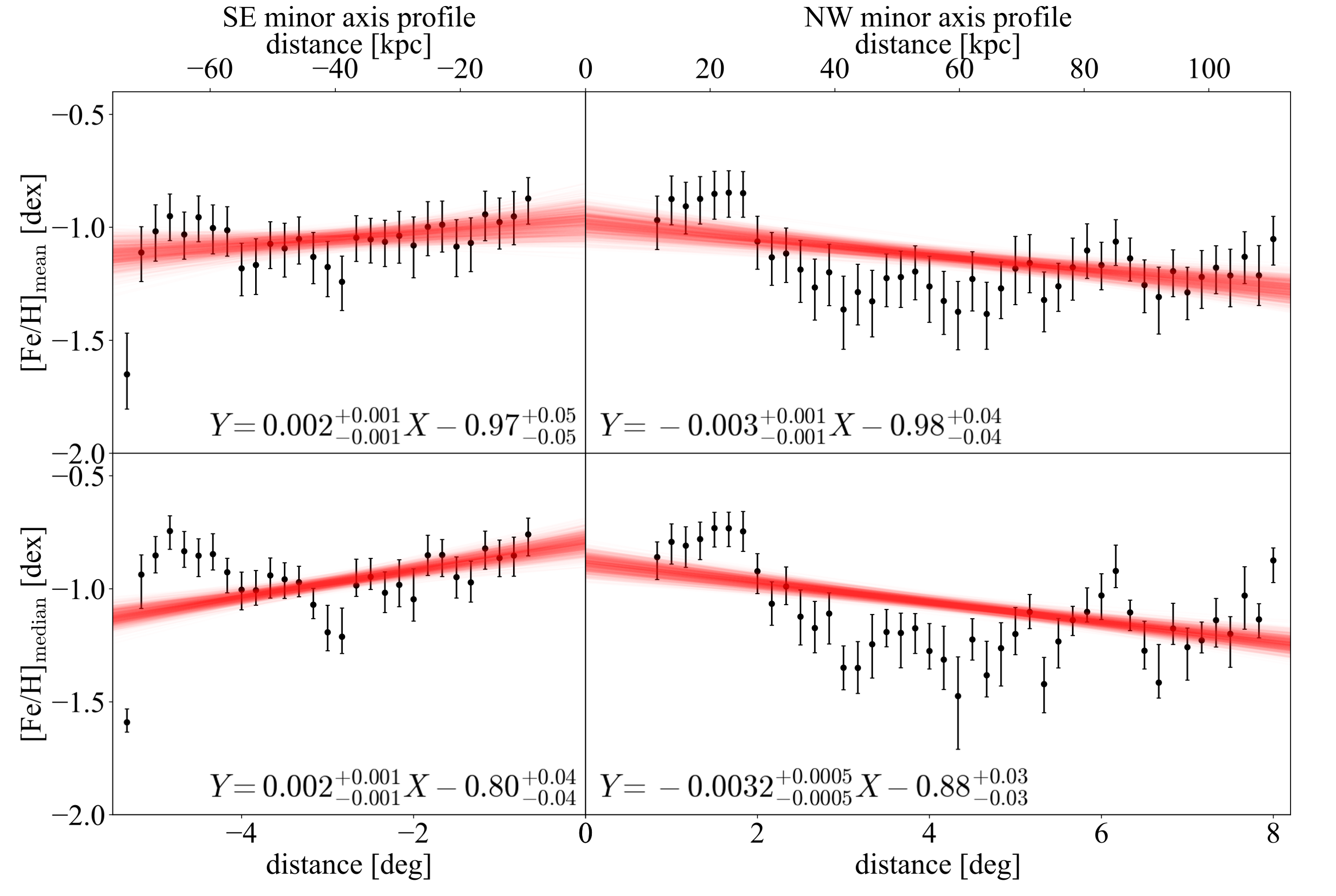}
 \caption{The metallicity profile along the minor-axis of the M31 disc. The left figure is the southeastern minor axis profile and the right figure corresponds to the northwestern minor axis profile. The plots are the same as in Figure \ref{figure:MD_RadialProfile}.}
 \label{figure:MD_MinorProfile}
\end{figure*}

Figure \ref{figure:MD_RadialProfile} shows the radial metallicity profile of the M31 halo and Figure \ref{figure:MD_MinorProfile} is the SE and NW minor axis profiles. In each panel, the black dots are the mean/median metallicity values in each binned region. We performed MCMC linear regression on these data using emcee with $chains=60, step=110000$. The solid red lines in Figure \ref{figure:MD_RadialProfile} and \ref{figure:MD_MinorProfile} are 1000 linear regression lines randomly sampled from the posterior distribution. The slope of the regression line for the radial profile for mean (median) metallicity is $-0.004^{+0.001}_{-0.001}$ dex kpc$^{-1}$ ($-0.0044^{+0.0004}_{-0.0004}$ dex kpc$^{-1}$), indicating that the halo of M31 has a metallicity gradient.

The SE minor axis profile for has a slope in the mean/median metallicity of $0.002^{+0.001}_{-0.001}$ dex kpc$^{-1}$ (the reason why the slope value is a positive value is that the distance from the centre of M31 is set to a negative value), and we can see the metallicity gradient in this profile. Moreover, the NW minor axis profile has a slope in the mean (median) metallicity of $-0.003^{+0.001}_{-0.001}$ dex kpc$^{-1}$ ($-0.0032^{+0.0005}_{-0.0005}$ dex kpc$^{-1}$). Therefore, in the present analysis, we confirm the existence of a metallicity gradient in both radial and minor-axis directions. It should be noted that substructures of M31 are included in the analysis, so the results reflect both smooth and substructure components in M31. It is difficult to separate the M31 substructures from the smooth halo using only the observational data in this study. Further kinematic information from future spectroscopic observations, like PFS, will be important, so that we can explore the metallicity properties with respect to each of the components.

\subsubsection{Surface brightness profile of all resolved RGB stars}\label{subsubsection:SBprofile}
\begin{figure}
 \includegraphics[width=\columnwidth]
 {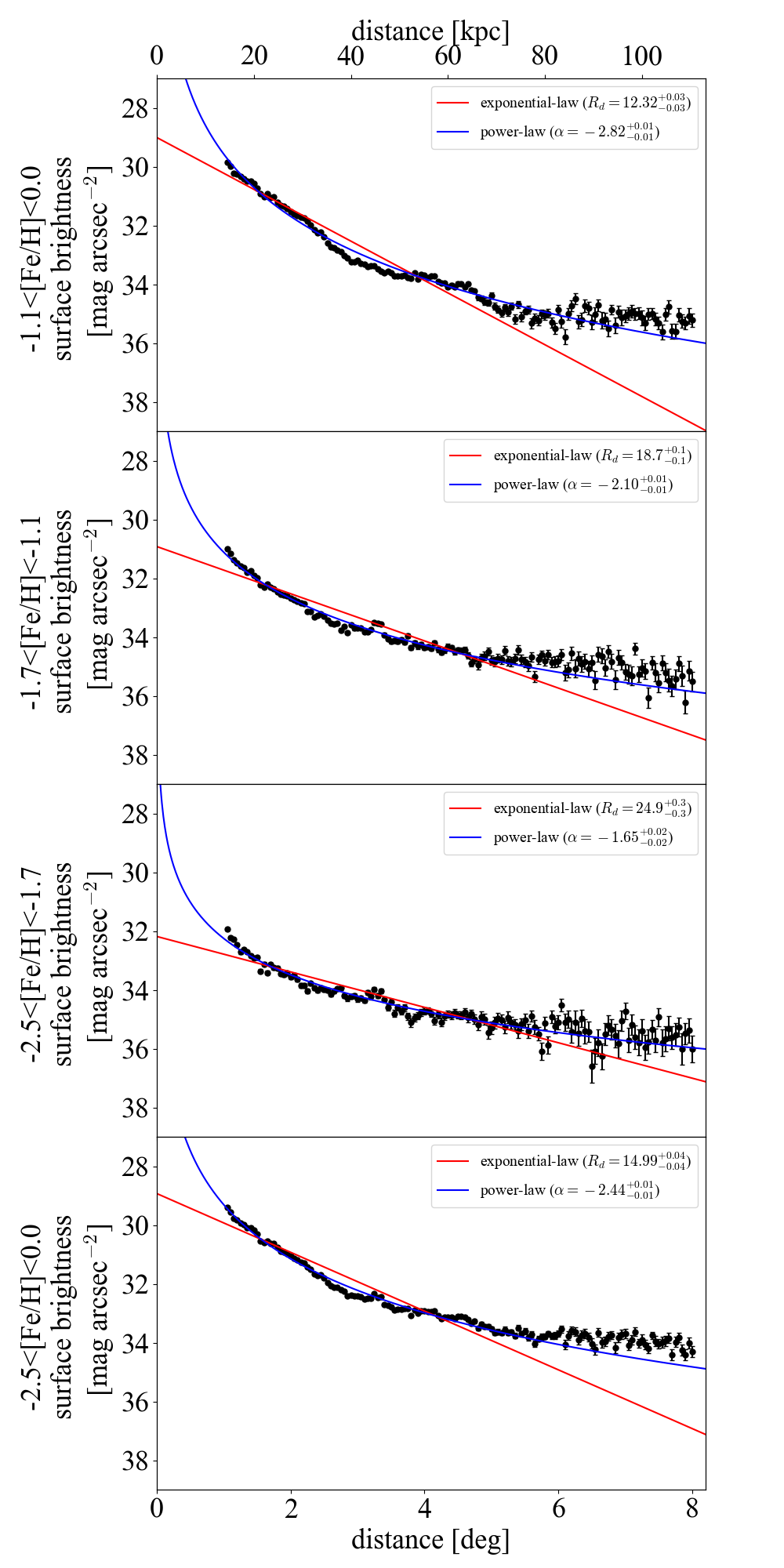}
 \caption{Radial surface brightness profile with the metallicity range of $-1.1<[\mathrm{Fe/H}]<0.0,~-1.7<[\mathrm{Fe/H}]<-1.1,~-2.5<[\mathrm{Fe/H}]<-1.7,~-2.5<[\mathrm{Fe/H}]<0.0$ from top to bottom. The blue dots are observed data, and the solid and dotted lines are the results of fitting to the exponential and power-law models.}
 \label{figure:SBRadialProfile}
\end{figure}

\begin{figure*}
     \includegraphics[width=2\columnwidth]
 {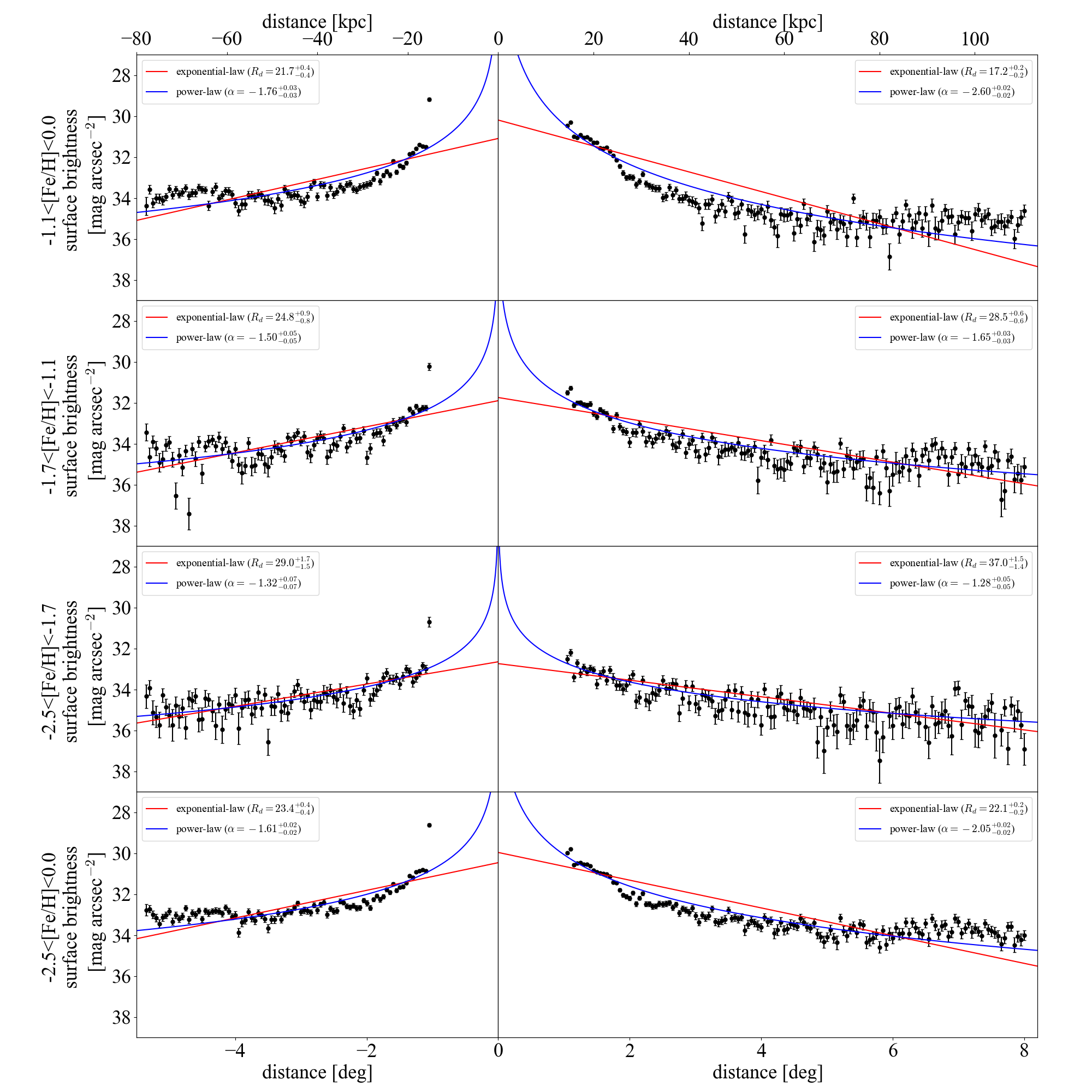}
 \caption{Surface brightness profile along the southeast (left) and northwest (right) direction of the M31 minor axis. The plots are the same as in Figure \ref{figure:SBRadialProfile}.}
 \label{figure:SBMinorProfile}
\end{figure*}

We construct the radial and two minor-axis surface brightness profiles using the NRGB stars defined in Section \ref{subsection:selection_RGB}. It should be noted that the profile we build in this study is ``the surface brightness of all resolved RGB stars'', but hereafter we call it simply as ``the surface brightness profile''. The surface brightness profile is constructed by adding the fluxes of the individual halo stars as follows. First, the M31 halo is divided into the same small regions as the metallicity profiles (Section \ref{subsubsection:MD}). Second, we divide the objects in each subregion by photometric metallicity (Section \ref{subsubsection:MD}). The datasets are divided according to \citep{ibata2014}: $-2.5 < [\mathrm{Fe/H}] < -1.~-1.7 < [\mathrm{Fe/H}] < -1.1,~-1.1 < [\mathrm{Fe/H}] < 0.0$. Third, the fluxes of the sources within each divided region are added and their V-band magnitude is derived according to the photometric transformation of \cite{ibata2007}. Finally, the surface brightness profile is constructed by dividing the sum of the added fluxes for each region by the area of the region. The constructed radial profile is shown in Figure \ref{figure:SBRadialProfile}, and minor axis profiles are shown in Figure \ref{figure:SBMinorProfile}. In addition to the surface brightness profiles for each metallicity range, the profiles for all metallicity ranges ($-2.5 < [\mathrm{Fe/H}] < 0.0$) are also included in Figures \ref{figure:SBRadialProfile} and \ref{figure:SBMinorProfile} for reference. 

In the left panel of Figure \ref{figure:SBMinorProfile}, the surface brightness tends to decrease as the distance from the centre increases, while the profiles for distances $\sim -3 \degr$ (-40 kpc) and $\sim -4 \degr$ (-55 kpc) show a hump in the surface brightness with $-1.7 < [\mathrm{Fe/H}] < -1.1$. This reflects the presence of Stream C and Stream D in this region. Similarly, for $-2.5 < [\mathrm{Fe/H}] < -1.7$ and $-1.7 < [\mathrm{Fe/H}] < -1.1$ in the right panel of Figure \ref{figure:SBMinorProfile}, the surface brightness profile is enhanced due to the presence of the low-metallicity NW Stream at the radius of $\sim 7 \degr$ (95 kpc) from the centre. The surface brightness also increases for $-1.1 < [\mathrm{Fe/H}] < 0$ at $\sim 7 \degr$ (95 kpc) in the right panel, which is due to a remnant of foreground stars in the Galaxy in this region. 

\begin{table*}
 \caption{The results of the surface brightness profiles.}
 \label{table:SB}
 \begin{tabular}{@{}l@{\hspace*{15pt}}l@{\hspace*{15pt}}l@{\hspace*{15pt}}l@{\hspace*{15pt}}l@{\hspace*{15pt}}l@{}}
      \hline
      Studies & $R_{min}$ & $R_{max}$ & power-law index & data type & memo\\
      & [kpc] & [kpc] & (scale length [kpc]) & (model) & \\
      \hline
      \citet{guhathakurta2005} & 30 & 165 & $-2.3$ & Spectroscopic & the south part of the M31 halo\\
      \citet{irwin2005} & 20 & 55 & $-2.3$ & Photometric & South-Eastern minor axis\\
      &  &  & (14) & (Exponential) & \\
      \cite{ibata2007} & 30-35 & 90-130 & $-1.9\pm0.12$ & Photometric & South-Eastern minor axis\\
      &  &  & (31.6$\pm$1.0) & (Exponential) & \\
      &  &  & (54.6$\pm$1.3) & (Hernquist) & \\
      & 90 & 130 & (45.1$\pm$6.0) & (Exponential) & \\
      &  &  & (53.1$\pm$3.5) & (Hernquist) & \\
      \citet{tanaka2010} & 20 & 100 & $-1.75\pm0.13$ & Photometric & South-Eastern minor axis\\
      &  &  & (22.4$\pm$2.3) & (Exponential) & \\
      &  &  & (31.7$\pm$6.7) & (Hernquist) & \\
      & 20 & 100 & $-2.17\pm0.15$ & & North-Western minor axis\\
      &  &  & (18.8$\pm$1.8) & (Exponential) & \\
      &  &  & (17.1$\pm$4.7) & (Hernquist) & \\
      \citet{courteau2011} & 0 & 165 & $-2.5\pm0.2$ & Photometric and Spectroscopic & South-Eastern minor axis\\
      \citet{williams2012} & 2 & 35 & $-2.6^{+0.3}_{-0.2}$ & Photometric & Using the horizontal branch stars\\
      \citet{gilbert2012} & 9 & 176 & $-2.2\pm0.2$ & Spectroscopic & statistically removed substructures\\
      & 35 & 176 & $-1.9\pm0.4$ & & all M31 stars\\
      & 9 & 90 & $-2.2\pm0.3$ &  & statistically removed substructures\\
      \cite{ibata2014} & 27 & 150 & $-2.08\pm0.02(\pm0.12)$ & Photometric & statistically removed substructures\\
      &  & &  &  & $-2.5<[\mathrm{Fe/H}]<-1.7$\\
      & 27 & 150 & $-2.13\pm0.02(\pm0.12)$ & Photometric & statistically removed substructures\\
      &  & &  &  & $-1.7<[\mathrm{Fe/H}]<-1.1$\\
      & 27 & 150 & $-2.66\pm0.02(\pm0.19)$ & Photometric & statistically removed substructures\\
      &  & &  &  & $-1.1<[\mathrm{Fe/H}]<0.0$\\
      This study & 14 & 110 & $-1.65^{+0.02}_{-0.02}$  & & the M31 halo entire region (radial profile)\\
      &  &  & ($24.9^{+0.3}_{-0.3}$) & (Exponential) & $-2.5<[\mathrm{Fe/H}]<-1.7$\\
      &  &  & $-2.10^{+0.01}_{-0.01}$ & & $-1.7<[\mathrm{Fe/H}]<-1.1$\\
      &  &  & ($18.7^{+0.1}_{-0.1}$) & (Exponential) & \\
      &  &  & $-2.82^{+0.01}_{-0.01}$ & & $-1.1<[\mathrm{Fe/H}]<0$\\
      &  &  & ($12.32^{+0.03}_{-0.03}$) & (Exponential) & \\
      &  &  & $-2.44^{+0.01}_{-0.01}$ & & $-2.5<[\mathrm{Fe/H}]<0$\\
      &  &  & ($14.99^{+0.04}_{-0.04}$) & (Exponential) & \\
      & 14 & 110 & $-1.32^{+0.07}_{-0.07}$  & & South-Eastern minor axis\\
      &  &  & ($29.0^{+1.7}_{-1.5}$) & (Exponential) & $-2.5<[\mathrm{Fe/H}]<-1.7$\\
      &  &  & $-1.50^{+0.05}_{-0.05}$ & & $-1.7<[\mathrm{Fe/H}]<-1.7$\\
      &  &  & ($24.8^{+0.9}_{-0.8}$) & (Exponential) & \\
      &  &  & $-1.76^{+0.03}_{-0.03}$ & & $-1.1<[\mathrm{Fe/H}]<0$\\
      &  &  & ($21.7^{+0.4}_{-0.4}$) & (Exponential) & \\
      &  &  & $-1.61^{+0.02}_{-0.02}$ & & $-2.5<[\mathrm{Fe/H}]<0$\\
      &  &  & ($23.4^{+0.4}_{-0.4}$) & (Exponential) & \\
      & 14 & 110 & $-1.28^{+0.05}_{-0.05}$ & & North-Western minor axis\\
      &  &  & ($37.0^{+1.5}_{-2.4}$) & (Exponential) & $-2.5<[\mathrm{Fe/H}]<-1.7$\\
      &  &  & $-1.65^{+0.03}_{-0.03}$ &  & $-1.7<[\mathrm{Fe/H}]<-1.1$\\
      &  &  & ($28.5^{+0.6}_{-0.6}$) & (Exponential) & \\
      &  &  & $-2.60^{+0.02}_{-0.02}$ & & $-1.1<[\mathrm{Fe/H}]<0$\\
      &  &  & ($17.2^{+0.2}_{-0.2}$) & (Exponential) & \\
      &  &  & $-2.05^{+0.02}_{-0.02}$ & & $-2.5<[\mathrm{Fe/H}]<0$\\
      &  &  & ($22.1^{+0.2}_{-0.2}$) & (Exponential) & \\
      \hline
      \end{tabular}
\end{table*}

We then perform a model fitting to each of these profiles. It has been shown that the surface brightness in the M31 halo can be expressed as a power law \citep[$\Sigma(R) \propto R^{-\al}$ or an exponential law ($\Sigma(R) \propto \exp{(-R/R_s)}$, where $R_s$ is the scale length. e.g.,][]{guhathakurta2005,irwin2005,ibata2007,tanaka2010}. Figures \ref{figure:SBRadialProfile} and \ref{figure:SBMinorProfile} show the power-law and exponential-law models fitted to the observed data with dotted and solid lines, respectively. Table \ref{table:SB} shows the results of the fitting and previous studies. For both minor axis and radial profiles, we are able to fit power-law models with power indices ranging from $\sim -1.3$ to $\sim -3.0$. In \citet{pritchet1994}, the M31 halo is represented by a single de Vaucouleurs's law ($\ln I \propto -R^{1/4}$) between 0.2 kpc and 20 kpc. The results of this study show that the power index is $\sim -1.3$ to $\sim -3.0$, which is different from the structure of the inner halo represented by de Vaucouleurs's law with a power index of $1/4$. This may reflect the fact that the M31 halo is not a single structure. 

Our model fitting using the exponential law results in a scale length range from $\sim 10$ to $\sim 40$ kpc. Table 5 shows that the scale lengths in this study are consistent with those of \citet{tanaka2010}. However, in Figures \ref{figure:SBRadialProfile}, and \ref{figure:SBMinorProfile}, it can be seen that the exponential law reproduces the surface brightness near the centre, but does not reproduce the M31 halo in the outer region ($> 80$ kpc). Therefore, the power law is more suitable than the exponential law for reproducing the entire M31 halo for the radial range of this study. 

The higher the metallicities of the samples in the profiles, the smaller the $\alpha$ in the power law and the smaller the scale length in the exponential law, indicating a steeper distribution for the metal-rich objects. This result suggests the fact that more massive (metal-rich) mergers sink further into the halo and the outer halo consists of mostly minor (metal-poor) mergers. In addition, this also is consistent with hierarchical $\Lambda$-CDM paradigm where the stellar halo is predicted to have been mainly formed by accretion events \citeg{johnston2008,sharma2011}. In these studies, the metal-rich halos are formed by massive accretion events. More massive the accreting system is, the stronger is the dynamical friction, so massive accretion events dominate the inner halo and will have a sharper brightness profile. In contrast, the outer halo is dominated by a lot of small accretion events and should have a shallower brightness profile. Instead of this scenario, the origin of these metallicity gradients might be the gradients in the accreted dwarf galaxies. This result is also confirmed by the metallicity estimation results in Section \ref{subsubsection:MD}.

The power indices of the radial and minor axis profiles with low metallicity (e.g., $-2.5<[\mathrm{Fe/H}]<-1.7$) in this study do not agree with those of \cite{ibata2014}. In particular, the profiles with $-2.5<[\mathrm{Fe/H}]<-1.7$ in this study tend to be shallower than that in Ibata et al. This may be due to the inclusion of substructures (e.g., NW Stream) in the present profiles, resulting in higher surface brightness in the outer region and shallower gradients in the profiles. In fact, \cite{ibata2014} constructed the surface brightness profile by masking the substructure after statistically removing the contamination sources beforehand, suggesting that masking the substructure in the same way will result in a steeper gradient in our surface brightness profiles.

\section{Discussion}\label{section:discussion}

\subsection{Accuracy of the M31 RGB stars selection}\label{subsection:accuracy}
\begin{figure}
 \includegraphics[width=\columnwidth]
 {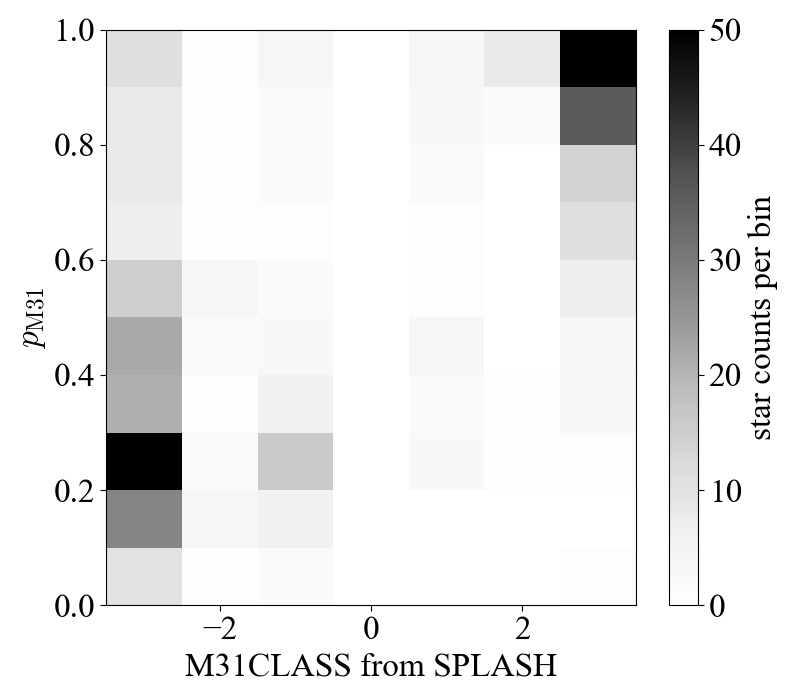}
 \caption{Comparison of $p_{\mathrm{M31}}$ calculated in this study with the label, "M31CLASS", which is the M31 RGB star as determined by the SPLASH data. In "M31CLASS", the more positive (negative) the value, the more likely it is that a star is an M31 RGB star (foreground dwarf star).}
 \label{figure:accuracy}
\end{figure}

In Section \ref{subsection:selection_RGB}, we calculate the probability $p_{\mathrm{M31}}$ that the stars are likely to be RGB stars in M31 based on \textit{NB515} and Galactic latitude information. In this subsection, we discuss its accuracy of $p_{\mathrm{M31}}$ by comparing with available spectroscopic information of the stars.

Spectroscopic observations were conducted targeting the inner halo of the M31 \citeg{ibata2004,chapman2008,collins2009}. Besides these pencil beam observations, some large-scale spectroscopic surveys were also conducted recently \citeg{gilbert2009,escala2022a,dey2023}. To compare our results, we use the SPLASH data \citep{gilbert2018}, which has many spectroscopically identified stars and is labeled as whether each star is a giant or dwarf star. Based on the line-of-sight velocity, the strength of the Na I doublet absorption line, colour, magnitude, and metallicity, the SPLASH data are attached the labels to individual stars that are likely to be a dwarf or RGB star (M31 CLASS). In this label, the positive value ($+1$, $+2$) means that the object is likely to be an M31 RGB star, and the negative value ($-1$, $-2$) means that it is another source such as the foreground dwarf star. Therefore, we compare this label with $p_{\mathrm{M31}}$. Figure \ref{figure:accuracy} shows a two-dimensional histogram with M31 RGB probability on the vertical axis, and SPLASH label on the horizontal axis. In this figure, the plots in the upper right and lower left are those considered to be M31 members/non-members in both the HSC and SPLASH catalogues. To obtain the accuracy rate of our RGB selection, we evaluate it from the following fraction: the number of objects determined to be RGB/MS in HSC and SPLASH divided by the number of objects cross-matched in HSC and SPLASH. 

\begin{align}
&\mathrm{accuracy~rate} = \notag\\
&\f{n(\mathrm{stars~that~have~the~same~RGB~(dwarf)~label~in~both~catalogues})}{n(\textrm{cross-matched~stars})}
\end{align}
, where the RGB (dwarf) stars in the HSC catalogue are labeled as the stars with $p_{\mathrm{M31}} > 0.5$ ($p_{\mathrm{M31}} <0.5$), and the RGB (dwarf) stars in the SPLASH catalogue are labeled as the stars with $\mathrm{M31CLASS} > 0$ ($\mathrm{M31CLASS} < 0$). Based on this formula, we calculate the accuracy rate of our RGB selection to be 90.2\%, thereby demonstrating that \textit{NB515} is a powerful tool for separating RGB and dwarf stars.

\subsection{Comparison with Simulation}\label{subsection:simulation}

Simulation studies for reproducing these M31 substructures have been conducted by several researchers \citep[e.g.,][]{fardal2007,mori2008,hammer2018}. In this subsection, we compare our analysis for GSS and NW Stream with some simulation results. 

Many simulations have been performed to reproduce the orbits of the GSS. In particular, two hypotheses for the formation scenario of the GSS have been discussed theoretically: the minor merger hypothesis \citep{fardal2007,mori2008,kirihara2017a}, which postulates that small galaxies collided about 0.6 Gyr ago, and the major merger hypothesis \citep{hammer2018}, which postulates that massive galaxies collided with each other a few Gyr ago. Since the distance values estimated in this study are statistically consistent with the previous studies \citep{mcconnachie2003,conn2016}, it is difficult to constrain the two hypotheses using the line-of-sight distance. However, two of the three substructures (Metal-Poor Cloud and SE Stream) detected in this study might be used to constrain the two hypotheses. For the  SE Stream, similar structures have been identified in \citet{miki2016} and \citet{hammer2018}, so this structure is expected to form regardless of the mass of the progenitor. As for the Metal-Poor Cloud, its existence was predicted in \citet{kirihara2017a}. The details of the origin and nature of these substructures are discussed in Section \ref{subsection:newsubstructures}, but it is suggested that two of the three structures detected in this study may need to be minor merger in order to reproduce simultaneously in the GSS formation event. However, the possibility that these structures are independent of the GSS cannot be rejected, and further kinetic and scientific information is needed.

For the NW Stream, \citet{kirihara2017} proposed two groups of orbits (Case A and B) that reproduce the NW Stream (see their Figure 4). In Case A, the progenitor accretes from the northwest of the M31 halo toward the centre, resulting in the distance for most parts of the NW Stream being distributed farther than that of M31. In this case, the stream shows a distance gradient where the stream is distributed far in the line-of-sight direction where $\eta$ is large and closer where $\eta$ is small. In contrast, in Case B, the progenitor accretes from the eastern part of the M31 halo toward the centre, and the distance of the stream is distributed entirely in front of M31. Our analysis here confirms that the NW Stream is distributed in the front of M31 where $\eta$ is small, and in the back where $\eta$ is large. Therefore, the trend of the distance gradient which is estimated by this study is consistent with that in Case A, although the slope of the distance gradient is slightly different. Also, in Case B, its distance distribution shows that the stream locates at the front of M31. On the other hand, in Case A, most part of the stream locates behind M31, and the stream near the centre of M31 locates at the front of M31. Therefore, the result of distance estimation in this study supports the orbital model in Case A. In any case, the photometric distance results in this study and the line-of-sight velocity data obtained by Subaru/PFS in future will allow us to place limits on stream formation events.

\subsection{New Substructures}\label{subsection:newsubstructures}

In this paper, we have reported on three new substructures (Metal-Rich Cloud, Metal-Poor Cloud, and SE Stream) in the M31 halo (see \ref{subsubsection:spatial}). In this section, we discuss the origin of these substructures, accompanied by the results of the distance estimation.
% About Metal-Poor Cloud
\begin{figure*}
\includegraphics[width=2\columnwidth]{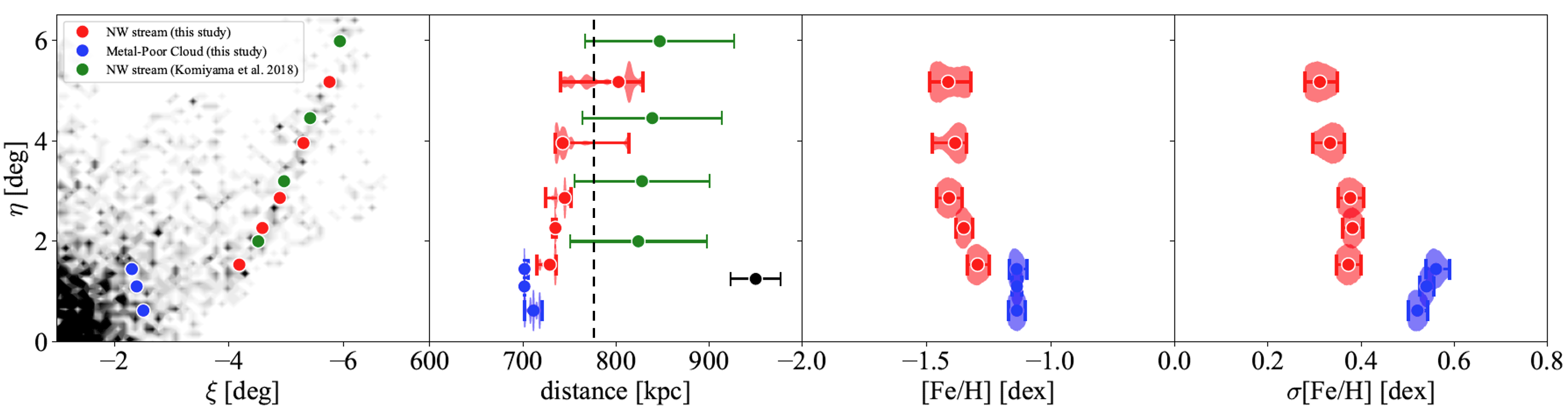} 
\caption{The same as in Figure \ref{figure:distance_GSS} but for the results of the distance estimation for the Metal-Poor Cloud. In each panel, black dots show the results of the Metal-Poor Cloud, and red and green dots correspond to the results of the NW Stream in this study and \citet{komiyama2018a}, respectively.}
\label{figure:distance_MPC}
\end{figure*}

Figure \ref{figure:distance_MPC} shows the distance distribution for the Metal-Poor Cloud, suggesting that this structure is distributed in front of M31. Previous PAndAS studies identified the stars that appear to be composed of the Metal-Poor Cloud \citep[e.g., Figure 9 in ][]{ibata2014}, but these stars are treated as a smooth halo in these studies. If this structure is the component of the fully dynamically relaxed smooth halo, the results of the distance estimation should have values around the M31 distance ($\sim 785$ kpc), given that the smooth halo is a spherically symmetric structure. However, this structure is located in front of M31, and the stellar populations are concentrated in metal-poor ranges, so we conclude that this structure is not a dynamically relaxed structure and we consider that this cloud is one of the new substructures in the M31 stellar halo.

%% The Origin of Metal-Poor Clouds
Since this region is adjacent to the Western Shell and the NW Stream, there are three possible origins for this substructure: (1) an origin from the accretion event with the GSS, (2) the NW Stream, or (3) an independent accretion origin. As is evident from Figure \ref{figure:distance_MPC}, the photometric metallicity of this substructure is in almost agreement with the NW Stream, and also the distance distribution appears continuous from the NW Stream. However, if we look at the shape of the spatial distribution of the Metal-Poor Cloud, this overdensity shows a broad structure, unlike the thin shape of the NW Stream. Instead, the Metal-Poor Cloud has a smooth shape, which appears to be continuous from the Western Shell. Other than the spatial distribution, the standard deviations of photometric metallicities of this overdensity are larger than those of the NW Stream. Moreover, the results in Section \ref{subsubsection:distances} also suggest that the GSS does not only have a metal-rich population but also contains a metal-poor component, so the metallicity distribution of the GSS includes the metallicity range of the Metal-Poor Cloud. In addition to the observational evidence, according to the simulation study \citep{kirihara2017a}, a metal-poor substructure is predicted to be outside the Western Shell originating from the GSS accretion event \citep[called ``Outer Western Shell'' by][]{kirihara2017a}. Therefore, we think that it is highly likely that this structure was formed by the accretion event of the GSS. We note that it is yet uncertain from only current photometric data whether this substructure is indeed related to the formation of the GSS, the NW Stream, or individual accretion. To resolve this question, a much deeper and wider spectroscopic observation (e.g., Subaru/PFS) to measure the stellar kinematics will be needed to understand the origin of this halo structure in detail.

%% Metal-Rich Cloud
In addition to the Metal-Poor Cloud, we have also found the Metal-Rich Cloud at $(\xi,\eta) \sim (-1\degr,2\degr)$ in Figure \ref{figure:RGB_Substructure} with $-0.6<[\mathrm{Fe/H}]<-0.1$. This structure may originate from (1) the GSS accretion event, or (2) an independent accretion event. Similar to the Metal-Poor Cloud, the metallicity range of the GSS covers the metallicity of the Metal-Rich Cloud. This overdensity is also spatially spread out and distributed near the centre of M31, where the GSS is dominant. Therefore, we think that the Metal-Rich Cloud is associated with the GSS, as is the case for the Metal-Poor Cloud.

%% SE Stream
Finally, we have found the SE Stream which is distributed in the M31 inner halo. Many of the substructures in the M31 inner halo (e.g., the Western Shell, the North-Eastern Shell, and the South-Eastern Shell) are thought to have been formed when the GSS was formed. In Figure 3a of \citet{miki2016} which simulated the GSS formation, stream-like structures can be seen in addition to the GSS and shell structures. Although the SE Stream is distributed outside of the locations identified in Fig. 3a of \citet{miki2016}, such a stream-like structure may have been formed during the GSS accretion, according to this simulation. Furthermore, \citet{hammer2018} conducted the simulation for the formation of the GSS, and this simulation model \citep[see, "\texttt{\#255}" model shown in Figure 8]{hammer2018} has almost the same shaped substructure as that of the SE Stream near the centre of M31, although this simulation is a different formation scenario where the progenitor is massive. So this stream is possible to be one of the substructures associated with the GSS. Also, this stream has similar metallicity to the Stream C and D. This suggests that it may have formed further inward at the same time as the formation of the Stream C and D, and/or associated with these streams. In any case, it is not possible to determine whether the SE Stream is associated with other substructures using only the photometry in this study. Therefore, we expect that future spectroscopic observations measuring radial velocities and detailed chemical abundances of stars will allow us to elucidate the properties of this substructure, and its origin will be revealed by comparing it with more detailed simulation studies.

\section{Conclusions}\label{section:conclusions}
We have carried out a wide-area narrow-band imaging survey of the stellar halo in M31 using HSC on the Subaru Telescope. This survey covers about 50 deg$^2$ field consisting of 33 HSC pointings in the narrowband filter, \textit{NB515}. Combining the PAndAS/\textit{g-} \& \textit{i-}bands, we have constructed a method to statistically extract M31 RGB stars (NRGB stars) using the colour-colour diagram and their galactic latitude. Using this method, we have succeeded in deriving the most likely inner halo structure of M31 without suffering from as large a degree of contamination from foreground stellar populations.

The spatial distribution of NRGB stars shows prominent stream features, including already-known streams (e.g., GSS, Western Shell, NW Stream). We have also performed the distance and metallicity estimates for these substructures, and found that the properties of the GSS identified in this study are in good agreement with previous studies. Moreover, the NW Stream shows a steeper distance gradient than that derived in the previous study, suggesting that it was formed in an orbit closer to the Milky Way. For the Streams C and D, we have resolved the distance gradients of these streams thanks to the \textit{NB515}-selection and the distance and metallicity of these streams are consistent with each other. In addition to these existing structures, we have discovered three new substructures by taking advantage of \textit{NB515}. Estimating photometric metallicities and metallicity dispersions, all substructures can be the tidal debris of the accretion event, which constructed the GSS.

Finally, we have derived the global halo photometric metallicity distribution and surface brightness profile using \textit{NB515}-selected halo stars. The surface brightness of the metal-poor stellar population can be fitted to a shallower power-law profile with $\al = -1.65 \pm 0.02$ than that of the metal-rich population with $\al = -2.82 \pm 0.01$. In contrast to the relative smoothness of the halo density profile, its metallicity distribution appears to be spatially non-uniform with nonmonotonic trends with radius, suggesting that the halo population had insufficient time to dynamically homogenize the accreted populations.

The accuracy of our \textit{NB515}-based selection reaches 90\%, indicating that this narrow-band filter is a powerful tool for the analysis of existing outer halo data and deep photometry of the inner halo, which are currently under analysis (Ogami et al. in prep). In addition, it is also expected to provide ideal targets for future large-scale spectroscopic observations.

\section*{Acknowledgements}

We acknowledge support in part from MEXT Grant-in-Aid for Scientific Research (No.~JP18H05437 and JP21H05448 for M.C. No.~JP21K13909, and JP23H04009 for K.H.). MGL was supported by the National Research Foundation grant funded by the Korean Government (NRF-2019R1A2C2084019). This work was partially supported by Overseas Travel Fund for Students (2023) of Astronomical Science Program, The Graduate University for Advanced Studies, SOKENDAI. Data analysis was in part carried out on the Multi-wavelength Data Analysis System operated by the Astronomy Data Center (NAOJ/ADC) and the Large-scale data analysis system co-operated by the ADC and Subaru Telescope, NAOJ.  E.N.K.\ acknowledges support from NSF CAREER grant AST-2233781.  Based in part on data collected at Subaru Telescope and obtained from the SMOKA, which is operated by the NAOJ/ADC. This research used the facilities of the Canadian Astronomy Data Centre operated by the National Research Council of Canada with the support of the Canadian Space Agency. The Pan-STARRS1 Surveys (PS1) and the PS1 public science archive have been made possible through contributions by the Institute for Astronomy, the University of Hawaii, the Pan-STARRS Project Office, the Max-Planck Society and its participating institutes, the Max Planck Institute for Astronomy, Heidelberg and the Max Planck Institute for Extraterrestrial Physics, Garching, The Johns Hopkins University, Durham University, the University of Edinburgh, the Queen's University Belfast, the Harvard-Smithsonian Center for Astrophysics, the Las Cumbres Observatory Global Telescope Network Incorporated, the National Central University of Taiwan, the Space Telescope Science Institute, the National Aeronautics and Space Administration under Grant No. NNX08AR22G issued through the Planetary Science Division of the NASA Science Mission Directorate, the National Science Foundation Grant No. AST–1238877, the University of Maryland, Eotvos Lorand University (ELTE), the Los Alamos National Laboratory, and the Gordon and Betty Moore Foundation.

%%%%%%%%%%%%%%%%%%%%%%%%%%%%%%%%%%%%%%%%%%%%%%%%%%
\section*{Data Availability}
The HSC raw images are publicly available at www.https://smoka.nao.ac.jp. The PAndAS images and reduced photometry catalog are also available at https://www.cadc-ccda.hia-iha.nrc-cnrc.gc.ca/en/community/pandas/query.html. % Reduced photometry can be obtained from the lead author upon reasonable request.

% The inclusion of a Data Availability Statement is a requirement for articles published in MNRAS. Data Availability Statements provide a standardised format for readers to understand the availability of data underlying the research results described in the article. The statement may refer to original data generated in the course of the study or to third-party data analysed in the article. The statement should describe and provide means of access, where possible, by linking to the data or providing the required accession numbers for the relevant databases or DOIs.



\bibliographystyle{mnras}
\bibliography{M31} % if your bibtex file is called example.bib

%\begin{thebibliography}{99}
%\end{thebibliography}

%%%%%%%%%%%%%%%%%%%%%%%%%%%%%%%%%%%%%%%%%%%%%%%%%%

%%%%%%%%%%%%%%%%% APPENDICES %%%%%%%%%%%%%%%%%%%%%

\appendix
\section{The method of artificial star test}\label{section:completeness}
In this section, we describe how to calculate detection completeness using artificial stars in the hsciPipe (especially after version 6). In this study, we develop a Python code \texttt{injectStar.py} to insert artificial stars into coadd-images and calculate the detection completeness using this code and hscPipe. 

The general algorithm of the \texttt{injectStar.py} is almost the same as \citet{aihara2018a}. The algorithm of the \texttt{injectStar.py} consists of two steps. To insert artificial stars, the first step in the \texttt{injectStar.py} is to obtain the point spread function (PSF) from each stacking image which is $4000~\mathrm{pixels} \times 4000~\mathrm{pixels}$ region called \texttt{patch} in the hscPipe. Using these obtained PSF, the second step is to construct the artificial stars given the apparent magnitude and inject in each co-added \texttt{patch} image, every $60$ pixel by $60$ pixels. After these processes, we apply the hscPipe detection and photometry algorithm (\texttt{detectCoaddSources.py} and \texttt{multiBandDriver.py}) to the images with embedded artificial stars. The series of these processes are repeated in $0.25$ magnitude over a range of magnitudes $18 \leq \textit{NB515} \leq 28$ to obtain the detection completeness, the percentage of detected artificial stars, for all patches. 

\begin{figure}
  \includegraphics[width=\columnwidth]
  {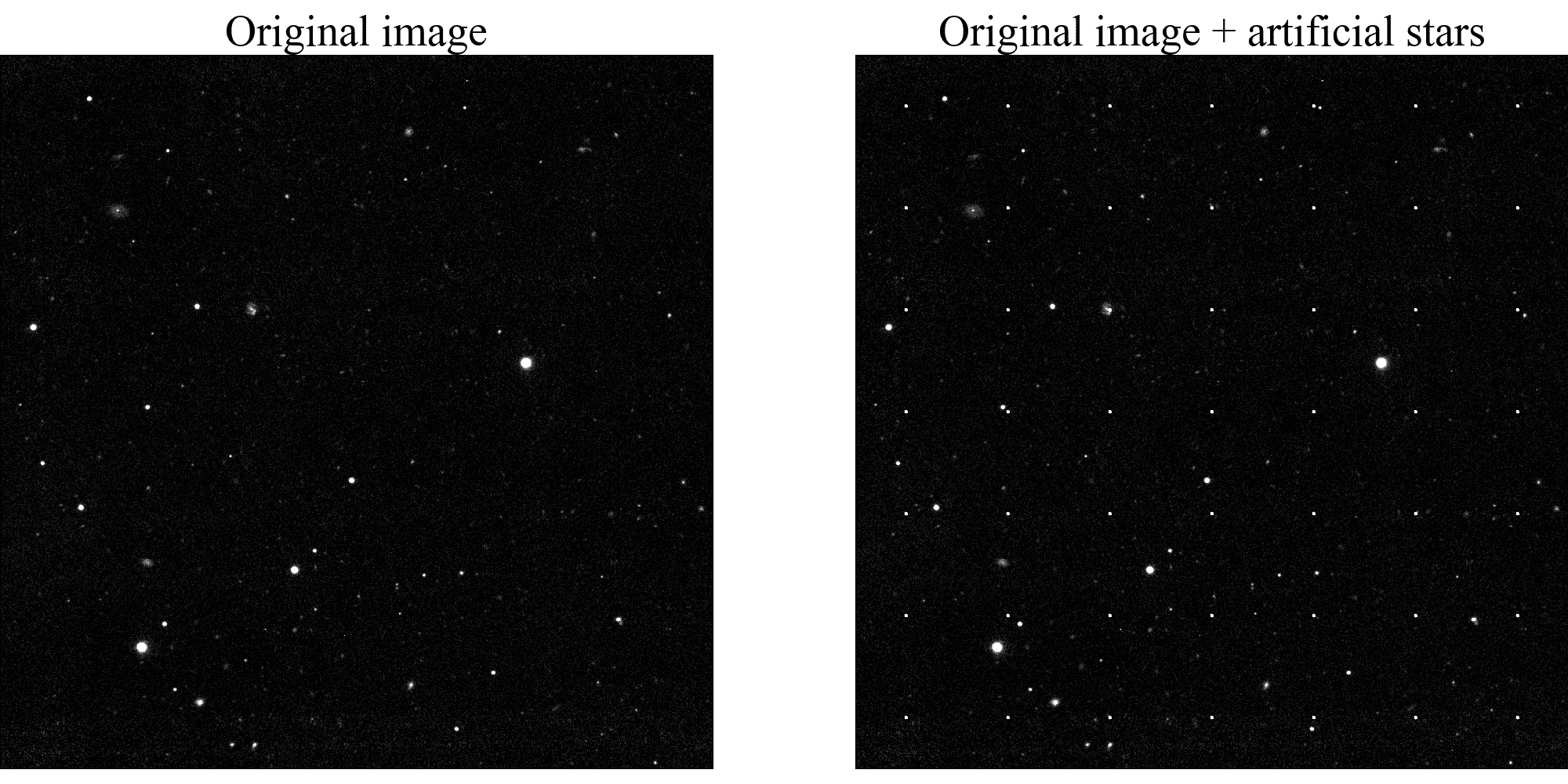}
 \caption{The \texttt{patch} images, which are before (left) and after (right) performing the \texttt{injectStar.py}. The magnitude of the embedded artificial star is 17 mag.}
 \label{figure:ArtificialStarTest}
\end{figure}

Figure \ref{figure:ArtificialStarTest} shows the \texttt{patch} images, which are before and after performing the \texttt{injectStar.py}. The magnitude of the embedded artificial star is 17 mag, and it can be seen that these stars are equally embedded in this \texttt{patch} image. 

%%%%%%%%%%%%%%%%%%%%%%%%%%%%%%%%%%%%%%%%%%%%%%%%%%

% Don't change these lines
\bsp	% typesetting comment
\label{lastpage}
\end{document}